\documentclass[aip,jcp,amsmath,reprint]{revtex4-1}
\usepackage{textcomp,graphicx,dcolumn,xspace}
\usepackage[T1]{fontenc}
\usepackage[ansinew]{inputenc}

\usepackage{hyperref}
\hypersetup{bookmarksnumbered=true,bookmarksopen,bookmarksopenlevel=1,
            colorlinks,allcolors=blue,urlcolor=black,
			bookmarksdepth=2,pdfpagemode=UseOutlines,pdfprintscaling=None,
            pdfauthor={Vincent Holten, Jeremy C. Palmer, Peter H. Poole, Pablo G. Debenedetti, and Mikhail A. Anisimov},
            pdftitle={Two-state thermodynamics of the ST2 model for supercooled water}}

\usepackage{mathptmx,afthd}
\DeclareMathAlphabet{\mathcal}{OMS}{cmsy}{m}{n}
\newcommand{\ea}{\textit{et al.}\xspace}
\newcommand{\figref}[1]{Fig.~\ref{#1}}
\newcommand{\ffigref}[1]{Figure~\ref{#1}}

\newcommand*{\eqsref}[2]{Eqs.~(\ref{#1}) and (\ref{#2})}

\newcommand*{\secref}[1]{Sec.~\ref{#1}}

\newcommand{\pypxl}[2]{{\partial{#1}/\partial{#2}}}
\newcommand{\pypxd}[3]{\left(\frac{\partial{#1}}{\partial{#2}}\right)_{#3}}
\newcommand{\di}{\mathrm{d}}

\newcommand*{\ts}[1]{_\text{#1}}
\newcommand*{\tu}[1]{^\text{#1}}
\newcommand*{\GnA}{G\tu{A}}

\newcommand*{\GnBA}{G\tu{BA}}
\newcommand*{\xe}{x\ts{e}}
\newcommand*{\Tc}{T\ts{c}}
\newcommand*{\Pc}{P\ts{c}}
\newcommand*{\rhoc}{\rho_\text{c}}
\newcommand*{\Dt}{\Delta \hat{T}}
\newcommand*{\Dp}{\Delta \hat{P}}
\newcommand*{\poolemodel}{ST2(I)\xspace}
\newcommand*{\palmermodel}{ST2(II)\xspace}

\newcommand*{\A}{\Delta \Gh}
\newcommand*{\Ax}{\Delta \Gh_\times}

\newcommand*{\xx}{x_\times}
\newcommand*{\fx}{\phi_{1,\times}}
\newcommand*{\Wx}{W_\times}
\newcommand{\dWxdx}{\pypxd{\Wx}{x}{W}}
\newcommand{\dxxdx}{\pypxd{\xx}{x}{W}}
\newcommand*{\Del}{\Delta}
\newcommand*{\GBA}{\Gh\tu{BA}}

\begin{document}
\renewcommand*{\eqref}[1]{Eq.~(\ref{#1})}
\title{Two-state thermodynamics of the ST2 model for supercooled water}
\author{Vincent Holten}
\affiliation{Institute for Physical Science and Technology and Department of Chemical and
Biomolecular Engineering,\\
University of Maryland, College Park, Maryland 20742, USA}
\author{Jeremy C. Palmer}
\affiliation{Department of Chemical and Biological Engineering, Princeton University,
Princeton,\\
New Jersey 08544, USA}
\author{Peter H. Poole}
\affiliation{Department of Physics, St. Francis Xavier University, Antigonish, Nova
Scotia B2G 2W5, Canada}
\author{Pablo G. Debenedetti}
\affiliation{Department of Chemical and Biological Engineering, Princeton University,
Princeton,\\
New Jersey 08544, USA}
\author{Mikhail A. Anisimov}
\email[Author to whom correspondence should be addressed. Electronic mail: ]
{anisimov@umd.edu} \affiliation{Institute for Physical Science and Technology and
Department of Chemical and
Biomolecular Engineering,\\
University of Maryland, College Park, Maryland 20742, USA}
\date{\today}

\begin{abstract}
Thermodynamic properties of the ST2 model for supercooled liquid water exhibit anomalies
similar to those observed in real water. A possible explanation of these anomalies is the
existence of a metastable, liquid--liquid transition terminated by a critical point. This
phenomenon, whose possible existence in real water is the subject of much current
experimental work, has been unambiguously demonstrated for this particular model by most
recent simulations. In this work, we reproduce the anomalies of two versions of the ST2
model with an equation of state describing water as a non-ideal ``mixture'' of two
different types of local molecular order. We show that the liquid--liquid transition in
the ST2 water is energy-driven. This is in contrast to another popular model, mW, in
which non-ideality in mixing of two alternative local molecular orders is entropy-driven,
and is not sufficiently strong to induce a liquid--liquid transition.
\end{abstract}

\maketitle

\section{Introduction}
Liquid water is still poorly understood. Unlike ordinary substances, cold liquid water
(near the triple point and, especially, in the supercooled region) and high-temperature
liquid water behave as differently as if they consisted of different molecules.
Highly-compressible, low-dielectric-constant water at high temperatures is a good solvent
for hydrocarbons. On the low-temperature side of the phase diagram, water is an almost
incompressible, high-dielectric-constant liquid, a good solvent for electrolytes, and
exhibits numerous anomalous properties. The most famous anomaly is the maximum of the
density at 4~\textdegree C. Even more striking is the anomalous behavior of water's
thermodynamic response functions, such the heat
capacity,\cite{angell1973,angell1982,tombari1999,arc00}
compressibility,\cite{speedy1976,kanno1979,mishima2010} and thermal
expansivity,\cite{hare87,kanno1980,terminassian1981} which becomes significantly more
pronounced in the metastable supercooled state, suggesting a possible divergence at a
temperature just below the homogeneous ice nucleation limit. One of the scenarios
formulated to explain the anomalous behavior of water is the existence of a
liquid--liquid transition terminated by a liquid--liquid critical
point.\cite{poole1992,mishima1998review,stanley2000,stokely2010} The existence of the
liquid--liquid transition in water, the so-called liquid water
polyamorphism,\cite{tanaka2000EPL,mishima2010review} has remained a fascinating, but
highly debated, hypothesis. Indirect experimental evidence that supports this scenario
comes from the observation of kinks in the melting curves of metastable ice
polymorphs.\cite{mishima1998,mishima2000,mishima2011} More recently, it has been shown
that water possesses two glass transitions,\cite{amannwinkel2013} an important
observation that is consistent with the existence of two different forms of liquid water.
However, the liquid--liquid separation in bulk water has not yet been directly observed
in experiment, because of the challenge posed by rapid ice formation. The formation of
ice can, in principle, be avoided on the time scales accessible with simulations of
water-like models, allowing the properties of such models to be investigated under deeply
supercooled conditions, thus providing additional insights into the nature of water's
anomalies. However, the existence of a liquid--liquid transition in a given water-like
model is sensitive to the details of the intermolecular interactions. Furthermore,
numerical artifacts associated with the implementation of simulation techniques can mask
its presence.\cite{discussion2013A}

There are two popular models for water that demonstrate similar thermodynamic anomalies
in the supercooled region but strikingly different phase behavior, ST2 and mW. In the ST2
model, each water molecule is comprised of five interaction sites arranged in a
tetrahedral geometry.\cite{stillinger1974} The ST2 model qualitatively reproduces many of
the anomalous properties observed in real water. Its over-structured tetrahedral order
enables one to study the anomalies at higher temperatures than for other models. As a
result, ST2 has been used frequently in the computational investigations of water's
possible polyamorphism. Most simulations of the ST2 model have demonstrated the existence
of the liquid--liquid separation in the supercooled
region.\cite{liu2009,sciortino2011,liu2012,kesselring2012,kesselring2013,cuthbertsonpoole2011,poole2013,palmer2013}
Liu \ea\cite{liu2012} observed two distinct liquid basins in the reversible free energy
surface generated by computing, for a given temperature and pressure, histograms of
density and an order parameter that distinguishes crystalline from amorphous
configurations.\cite{steinhardt1983} Poole \ea\cite{poole2013} used a similar
computational approach to perform an analysis for ST2 water modeled with the
reaction-field treatment of the long-ranged electrostatics. The findings of Liu
\ea\cite{liu2012} and Poole \ea\cite{poole2013} are also consistent with recent molecular
dynamics simulations of ST2 water by Kesselring \ea,\cite{kesselring2012} which show
so-called \textquotedblleft phase flipping\textquotedblright\ between the high-density
liquid and low-density liquid on microsecond time scales. However, Limmer and
Chandler\cite{limmer2011,limmer2013} in their simulations of the ST2 model observed only
two minima in the free energy surface, one corresponding to a high-density liquid and the
other to a low-density ice. They argued that the liquid--liquid transition in ST2 water,
reported by other authors, was actually a liquid--crystal transition. The origin of the
differences between the result of Limmer and Chandler and virtually all other studies of
the ST2 model is still incompletely understood and is currently under
investigation.\cite{discussion2013A} We also note the recent work by English
\ea,\cite{english2013} who reported that the interface between the coexisting low- and
high-density liquids is not stable. This result is in contrast with ongoing explicit
interface simulation results in two of our groups, which will be the subject of a future
publication.

Another interesting model of water is the mW model devised by Molinero and
Moore.\cite{molinero2009} The mW model, suitable for fast computations, is a monatomic
model of water; it models the water molecule as a single atom with only short-range
interactions. The mW model imitates the anomalous behavior of cold and supercooled water,
including the density maximum and the strong increase of the heat capacity and
compressibility in the supercooled region.\cite{molinero2009,moore2011,limmer2011} Limmer
and Chandler\cite{limmer2011,limmer2013} have shown that the mW model does not exhibit
liquid--liquid separation in the range studied (from 0~MPa to 1000~MPa, down to 160~K).
Indeed, Moore and Molinero\cite{moore2011} demonstrated that in this model the
supercooled liquid can no longer be equilibrated before it crystallizes and there is no
sign of a liquid--liquid transition in the supercooled regime.

Despite the obvious difference with respect to the existence of the metastable
liquid--liquid phase transition, one remarkable feature is shared by both ST2 and mW
models,\cite{cuthbertsonpoole2011,moore2009} and also possibly by real
water:\cite{nilsson2012,taschin2013} the existence of two alternative, but
interconvertible, configurations (or ``states'') of local molecular order, namely a
high-density liquid structure and a low-density liquid structure. A bimodal inherent
structure in simulated TIP4P/2005 liquid water was also found by Wikfeldt
\ea\cite{wikfeldt2011} Competition between these configurations naturally explains the
density anomaly and other thermodynamic anomalies in cold water. In particular, if the
excess Gibbs energy of mixing of these two states is positive, the nonideality of the
``mixture'' can be sufficient to cause liquid--liquid separation, or, at least, to
significantly reduce the stability of the homogeneous liquid phase and consequently
generate the anomalies in the thermodynamic response functions. The best, so far,
description of all currently available experimental data on thermodynamic properties of
supercooled water is achieved with an equation of state based on the hypothesis of
non-ideal, nearly athermal mixing of the two structures in water.\cite{holtentwostate}
However, since experimental data are not yet available beyond the homogeneous ice
nucleation limit, the possibility of a liquid--liquid transition in real water must be
examined by indirect means. The transition line is obtained from the extrapolation of the
properties far away from the transition, thus making the location of the critical point
very uncertain. The existence of a bimodal distribution of molecular configurations in
water is supported by X-ray absorption and emission
spectroscopy\cite{nilsson2012,tokushima2008,wernet2004} and an investigation of
vibrational dynamics,\cite{taschin2013} and there is information on the fraction of water
molecules involved in each alternative structure.\cite{wernet2004,huang2009,xu2009}
Information on this fraction is also readily obtained from numerical or analytical
calculations for water models. The fractions of molecules involved in the high-density
structure and in the low-density structure at various temperatures and pressures were
computed and reported by Moore and Molinero\cite{moore2009} for the mW model, by
Cuthbertson and Poole\cite{cuthbertsonpoole2011} for the ST2 model, and by Wikfeldt
\ea\cite{wikfeldt2011} for the TIP4P/2005 model.

In this paper, we reproduce the anomalies of two versions of the ST2 model with an
equation of state describing water as a non-ideal \textquotedblleft
mixture\textquotedblright\ of two different types of local molecular order. We show that
the liquid--liquid transition in the ST2 water is mainly energy driven. This is in
contrast to the mW model, in which non-ideality in mixing of two alternative local orders
is entropy driven,\cite{holtenmW} and is not sufficiently strong to induce a
liquid--liquid transition.

\section{Thermodynamics of two states in liquid water}
\subsection{One liquid -- two structures}
Liquid--liquid phase separation in a pure substance can be explained if the substance is
viewed as a mixture of two interconvertible states or structures involving the same
molecules, whose ratio is controlled by thermodynamic equilibrium.\cite{bertrand2011} The
existence of two states does not necessarily mean that they will phase
separate.\cite{holtenmW,tanaka2013,discussion2013A} If these states form an ideal
solution, the liquid will remain homogeneous at any temperature or pressure. However, if
the solution is non-ideal, a positive excess Gibbs energy of mixing, $G\tu{E} = H\tu{E}-T
S\tu{E}$, could cause phase separation if the nonideality of mixing of the two states is
strong enough. If the excess Gibbs energy is associated with a heat of mixing $H\tu{E}$,
the separation is energy-driven. If the excess Gibbs energy is associated with an excess
entropy $S\tu{E}$, the separation is entropy-driven. The entropy-driven nature of such a
separation means that the two states would allow more possible statistical
configurations, and thus higher entropy, if they were unmixed.

Experiments\cite{nilsson2012,taschin2013} are consistent with the existence of a bimodal
distribution of molecular configurations in water. On a molecular level, Mishima and
Stanley\cite{mishima1998review} explained that if the intermolecular potential of a pure
fluid could exhibit two minima, the interplay between these minima may define the
critical temperature $\Tc$ and pressure $\Pc$ of liquid--liquid separation. Another
possibility is a double-step potential that depends on hydrogen-bond bending, as shown by
Tu \ea\cite{tu2012} A liquid--liquid transition was also demonstrated in the two-scale
spherically symmetric Jagla ramp model of anomalous liquids.\cite{xu2006}

The idea that water is a ``mixture'' of two different structures dates back to the 19th
century.\cite{whiting1884,roentgen} More recently, two-state models have become popular
to explain liquid polyamorphism.\cite{mcmillan2004,mcmillan2006,vedamuthu1994}
Ponyatovsky \ea\cite{ponyatovsky1998} and Moynihan\cite{moynihan1997} assumed that water
could be considered as a ``regular binary solution'' of two states, which implies that
the phase separation is driven by energy. They made an attempt to describe the
thermodynamic anomalies with this model, but the agreement with experimental data was
only qualitative.

Cuthbertson and Poole\cite{cuthbertsonpoole2011} applied the energy-driven version of the
two-state thermodynamics to describe the fraction of molecules in the high-density
structure of the ST2 model, which exhibits liquid--liquid separation. Holten
\ea\cite{holtenmW} explained and reproduced the thermodynamic anomalies of the mW model
with the same equation of state as was used in Ref.~\onlinecite{holtentwostate} to
describe real supercooled water. The direct computation of the fraction of molecules
involved in the low-density structure in the mW model was in agreement with the
prediction of the equation of state. However, in the mW model the athermal,
entropy-driven, non-ideality of mixing of the two alternative structures never becomes
strong enough to cause liquid--liquid phase separation. The situation in real water
remains less certain, but the best correlation of available experimental data for real
water favors an athermal, entropy-driven nonideality in mixing of such
configurations.\cite{holtentwostate}

The hypothesized existence of two liquid states in pure water can be globally viewed in
the context of polyamorphism,\cite{tanaka2000EPL,mishima2010review} a phenomenon that has
been experimentally observed or theoretically suggested in silicon, liquid phosphorus,
triphenyl phosphate, and in some other molecular-network-forming
substances.\cite{mcmillan2004,mcmillan2006,tanaka2013} Commonly, polyamorphism in such
systems is described as energy-driven. However, there is an ambiguity in terminology that
can be found in the literature,\cite{mcmillan2004,mcmillan2006} where the term ``density,
entropy-driven'' is used for a purely energy-driven phase separation as if the transition
were driven by both entropy and density.

The thermodynamic relation between the molar volume change $\Delta V$ and the latent heat
(enthalpy change) of phase transition $\Delta H = T\Delta S$ is given by the Clapeyron
equation
\begin{equation}
    \frac{\di P}{\di T}=\frac{\Delta S}{\Delta V}.
\end{equation}
Therefore, the relation between the volume/density ($\rho =1/V)$ change and the latent
heat/entropy change is controlled by the slope of the transition line in the $P$--$T$
plane. The Clapeyron equation itself does not provide an answer whether the
liquid--liquid transition in pure substances is energy-driven or entropy-driven. To
answer this question one should examine the source of non-ideality in the Gibbs energy.

\newcommand*{\Th}{\hat{T}}
\newcommand*{\Vh}{\hat{V}}
\newcommand*{\Ph}{\hat{P}}
\newcommand*{\Vc}{V_\text{c}}
\newcommand*{\Wt}{W_{\Th}}
\newcommand*{\Wp}{W_{\Ph}}
\newcommand*{\Wtt}{W_{\Th\Th}}
\newcommand*{\Wpp}{W_{\Ph\Ph}}
\newcommand*{\Wtp}{W_{\Th\Ph}}
\newcommand*{\Gh}{\hat{G}}
\newcommand*{\GA}{\Gh^{\text{A}}}
\newcommand*{\Sh}{\hat{S}}
\newcommand*{\kap}{\hat{\kappa}_T}
\newcommand*{\alp}{\hat{\alpha}_P}
\newcommand*{\Cph}{\hat{C}_P}

\subsection{Mean-field equation of state}\label{sec:meanfield}
We assume that liquid water at low temperatures can be described as a mixture of two
interconvertible states or structures, a high-density state A and a low-density state B.
The fraction of molecules in state B, denoted by $x$, is controlled by the `reaction'
\begin{equation}
\text{A}\rightleftharpoons \text{B}.  \label{eq:reaction}
\end{equation}
For the molar Gibbs energy $G$ of the two-state mixture, we adopt the following
expression:\cite{holtentwostate}
\begin{equation}  \label{eq:G}
G = \GnA + x\GnBA + RT\bigl[x \ln x + (1-x)\ln(1-x) + W x(1-x)\bigr],
\end{equation}
where $x$ is the mole fraction of state B, $\GnA$ is the Gibbs energy of pure state A,
$R$ is the molar gas constant, $T$ is the temperature, and $W$, the measure of the
nonideality of mixing, is a function of temperature and pressure.

The condition of chemical reaction equilibrium,
\begin{equation}\label{eq:dGdx}
\left( \frac{\partial G}{\partial {x}}\right) _{T,P}=0,
\end{equation}
defines the equilibrium fraction $x = \xe$.

The difference in Gibbs energy between the pure states $\GnBA \equiv G\tu{B} - \GnA$ is
related to the equilibrium constant $K$ of reaction~(\ref{eq:reaction})
by\cite{prigogine1954}
\begin{equation}
\ln K = -\frac{\GnBA}{RT}.
\end{equation}
The condition~(\ref{eq:dGdx}) implies
\begin{equation}  \label{eq:xeregular}
\ln K-\ln \frac{x}{1-x}- W (1-2x) =0.
\end{equation}
This equation must be solved numerically for the equilibrium fraction $x = \xe$. The
condition $\ln K = 0$ at $W>2$ defines the line of liquid--liquid transition between a
phase rich in structure A and a phase rich in structure B. The continuation of this line
($\ln K = 0$ at $W<2$) is known as the Widom
line.\cite{xu2005,fuentevilla2006,bertrand2011,holtenSCW,holtentwostate} The location of
the critical point ($\ln K = 0$ and $W=2$) is defined by
\begin{equation}\label{eq:critpoint}
    \biggl(\frac{\partial^2 G}{\partial x^2}\biggr)_{T,P} = 0, \qquad
    \biggl(\frac{\partial^3 G}{\partial x^3}\biggr)_{T,P} = 0.
\end{equation}
For the application to the ST2 model, we adopt a linear expression for $\ln K $ as the
simplest approximation,
\begin{equation}\label{eq:lnK}
\ln K = \lambda(\Dt + a\Dp),
\end{equation}
where
\begin{equation} \Dt =
\frac{T-\Tc}{\Tc},\quad \Dp = \frac{(P-\Pc)}{\rhoc R\Tc},
\end{equation}
with $\Tc$, $\Pc$, and $\rhoc$ the critical temperature, pressure, and molar density, and
the parameter $a$ is proportional to the slope of the $\ln K=0$ line in the phase
diagram,
\begin{equation}
    a = \frac{1}{\rhoc R}\frac{\di P}{\di T},
\end{equation}
and where $\lambda$ is related to the heat of reaction~(\ref{eq:reaction}),
\begin{equation}
    \lambda = \frac{\Delta H\tu{BA}\Tc}{R T^2}.
\end{equation}
We note that $\Delta H\tu{BA}$ is negative, and reaction~(\ref{eq:reaction}) is
exothermal.

In the theory of critical phenomena,\cite{fisher1983,anisimov2000inbook,behnejad2010} the
thermodynamic properties are expressed in terms of the scaling fields $h_1$ and $h_2$ and
the scaling densities $\phi_1$ and $\phi_2$, conjugate to $h_1$ and $h_2$. The scaling
density $\phi_1$, the order parameter, is associated with the low-density fraction as
\begin{equation}
    \phi_1 = (x - x\ts{c})/x\ts{c} = 2x - 1.
\end{equation}
The ordering field $h_1$, conjugate to the order parameter, and the second scaling field
$h_2$ are
\begin{align}
    h_1 &= \ln K,\label{eq:h1}\\
    h_2 &= 2 - W.\label{eq:h2}
\end{align}
The second scaling density $\phi_2$, conjugate to $h_2$, is in the mean-field
approximation
\begin{equation}
    \phi_2 = -\frac{1}{2}\phi_1^2.
\end{equation}
The strong susceptibility $\chi_1 = (\pypxl{\phi_1}{h_1})_{h_2}$ defines the
liquid--liquid stability limit (spinodal) as
\begin{equation}
    \chi_1^{-1} = \frac{1}{2RT} \biggl(\frac{\partial^2 G}{\partial x^2}\biggr)_{T,P} = 0,
\end{equation}
and is given in the mean-field approximation by
\begin{equation}\label{eq:chi1}
    \chi_1 = \biggl[\frac{1}{2x(1-x)} - W\biggr]^{-1}.
\end{equation}
The cross susceptibility $\chi_{12} = (\pypxl{\phi_1}{h_2})_{h_1} =
(\pypxl{\phi_2}{h_1})_{h_2}$ and the weak susceptibility $\chi_2 =
(\pypxl{\phi_2}{h_2})_{h_1}$ are given in the mean-field approximation by
\begin{align}
    \chi_{12} &= -\phi_1\chi_1,\\
    \chi_2 &= \phi_1^2\chi_1.
\end{align}
We now introduce dimensionless quantities for the temperature, pressure, molar Gibbs
energy, molar volume, and molar entropy, correspondingly,
\begin{gather}
\Th=\frac{T}{\Tc},\qquad \Ph=\frac{P}{\rhoc R\Tc}, \qquad \Gh=\frac{G}{R\Tc},\\
\Vh = \frac{\rhoc}{\rho}, \qquad \Sh=\frac{S}{R},
\end{gather}
and the dimensionless response functions, namely isothermal compressibility, expansivity,
and molar isobaric heat capacity,
\begin{equation}
    \kap = \rhoc R\Tc \kappa_T,\quad
    \alp = \Tc \alpha_P,\quad
    \Cph = C_P/R.
\end{equation}
The physical properties are given by (see Appendix~\ref{app:derivation})
\begin{align}
    \Vh &= -\frac{\Th}{2}\left[a \lambda(\phi_1+1) - \Wp(\phi_2+\tfrac{1}{2})\right] + \GA_{\Ph},\label{eq:Vh}\\
    \Sh &= -\frac{\Gh-\GA}{\Th} + \frac{\Th}{2}\left[\lambda(\phi_1+1) - \Wt(\phi_2 + \tfrac{1}{2})\right] - \GA_{\Th},\label{eq:Sh}
\end{align}
where the subscripts $\Th$ and $\Ph$ indicate partial derivatives with respect to these
quantities. With the assumption that $\Wpp=0$, we find for the response functions,
\begin{align}
    \kap\Vh &= \frac{\Th}{2}(a^2\lambda^2\chi_1 - 2a \lambda\Wp\chi_{12} + \Wp^2\chi_2) - \GA_{\Ph\Ph},\\
    \alp\Vh &= \frac{\Vh-\GA_{\Ph}}{\Th}
                - \frac{\Th\lambda}{2}(a \lambda\chi_1 - \Wp\chi_{12})\notag\\
                & + \frac{\Th}{2}\bigl[\Wtp(\phi_2 + \tfrac{1}{2}) + \Wt(a \lambda\chi_{12}-\Wp\chi_2)\bigr] + \GA_{\Th\Ph},\\
    \frac{\Cph}{\Th}
            &= \lambda(\phi_1 + 1) -\Wt(\phi_2 + \tfrac{1}{2}) + \Th\Bigl[\tfrac{1}{2}\lambda^2\chi_1 \\
                & - \lambda\Wt\chi_{12} + \tfrac{1}{2}\Wt^2\chi_2 - \frac{\Wtt}{2}(\phi_2 + \tfrac{1}{2})\Bigr]
                - \GA_{\Th\Th}.\label{eq:Cph}
\end{align}
The Gibbs energy $\GnA$ of the pure structure A defines the ``background'' of the
properties and is approximated as
\begin{equation}  \label{eq:background}
\GnA = R\Tc \sum_{m,n} c_{mn}(\Dt)^m (\Dp)^n,
\end{equation}
where $m$ and $n$ are integers and $c_{mn}$ are adjustable coefficients.

The nonideality factor $W$ may depend on temperature and pressure. If it does not depend
on the temperature, \eqref{eq:G} describes an athermal mixture (with non-ideal
entropy).\cite{prigogine1954} If $W$ is inversely proportional to the temperature, the
equation describes a regular mixture (with non-ideal enthalpy).\cite{prigogine1954} In a
more general case, both regular and athermal contributions can be present. We suggest the
following expression that contains both contributions:
\begin{equation}  \label{eq:W}
W = (1-\delta)(2 + \omega\ts{a} \Dp) + \frac{\delta(2+\omega\ts{r} \Dp)}{\Th},
\end{equation}
where $\delta$, the switching parameter between entropy-driven and energy-driven
nonidealities, has a value in the range of 0 to 1, and $\omega\ts{a}$ and $\omega\ts{r}$
are adjustable coefficients. For $\delta=0$, this yields a purely entropy-driven
nonideality of mixing, and for nonzero $\delta$ it yields an equation of state with both
entropy and energy contributions to the nonideality. A purely energy-driven nonideality
is obtained for $\delta=1$. For any value of $\delta$, the condition of $W=2$ at the
critical point is satisfied.

The Gibbs energy of mixing
\begin{align}\label{eq:deltaG}
    \A &= 2(\Gh-\GA-x\GBA) - \frac{W\Th}{2}\\
       &= 2\Th\biggl[x\ln x + (1-x)\ln(1-x) - \frac{W}{4}(1-2x)^2\biggr]
\end{align}
can be expanded around the critical point in powers of $\phi_1$ as
\begin{equation}\label{eq:phiexpansion}
      \frac{\A}{\Th}   \simeq \left(1-\frac{W}{2}\right)\phi_1^2 + \frac{1}{6}\phi_1^4 + \ldots,
\end{equation}
which can be compared with the Landau expansion\cite{bertrand2011}
\begin{equation}
    \frac{\A}{\Th} \simeq \frac{1}{2} h_2\phi_1^2 + \frac{1}{6}\phi_1^4 + \ldots
\end{equation}
For $\delta=0$ (entropy-driven nonideality), the second scaling field $h_2 =
-\omega\ts{a} \Dp$. For $\delta=1$ (energy-driven nonideality), a first-order expansion
in $\Dt$ and $\Dp$ yields
\begin{equation}
    h_2 = 2 - \frac{2+\omega\ts{r} \Dp}{\Th} \approx 2\Dt - \omega\ts{r}\Dp.
\end{equation}

\subsection{Accounting for critical fluctuations}\label{sec:crossover}
Accounting for critical order-parameter fluctuations, as predicted by scaling
theory,\cite{fisher1983,anisimov2000inbook,behnejad2010} in the vicinity of the critical
point is accomplished by a so-called crossover
procedure.\cite{anisimov2000inbook,behnejad2010,anisimov1992,kim2003b,holtentwostatesupplement}
To implement a crossover between mean-field and asymptotic scaling behavior, the
mean-field Gibbs energy is to be renormalized. This renormalization is carried out by
replacing the weak scaling field $h_2=2-W$ and the order parameter $\phi_1 = 2x-1$ by the
crossover variables $h_{2,\times} = 2-\Wx$ and $\fx = 2\xx-1$
as\cite{anisimov2000inbook,anisimov1992}
\begin{align}
    h_{2,\times} &= h_2 \,\mathcal{T}\mathcal{U}^{-1/2},\\
    \fx &= \phi_1 \,\mathcal{D}^{1/2}\mathcal{U}^{1/4},
\end{align}
where the rescaling functions $\mathcal{T}$, $\mathcal{U}$, and $\mathcal{D}$ will be
defined below. In addition, a kernel term $k$,
\begin{equation}
    k = -\tfrac{1}{2} c_t^2 h_2^2 \mathcal{K},
\end{equation}
responsible for the singularity in the weak susceptibility $\chi_2$, is to be added to
the renormalized Gibbs energy of mixing
$\Ax/\Th$.\cite{holtentwostatesupplement,edison1998,vanthof2001} Here $\mathcal{K}$ is
another rescaling function and $c_t$ is the kernel term amplitude. The detailed
description of the crossover theory can be found in
Ref.~\onlinecite{anisimov1992,kim2003b}. In this work, following
Ref.~\onlinecite{holtentwostatesupplement}, we use a simplified version of the crossover
procedure in which the parameter $c_t$ is defined as
\begin{equation}
    c_t = \tfrac{1}{2}(u^* \Lambda)^{1/2},
\end{equation}
where the fixed-point coupling constant of renormalization group theory $u^* \simeq
0.472$, and $\Lambda$ is a molecular cutoff.

The rescaling functions are defined as
\begin{align}
    \mathcal{T} &= Y^{(2\nu-1)/\Del},  &
    \mathcal{U} &= Y^{\nu/\Del},\\
    \mathcal{D} &= Y^{(\gamma-2\nu)/\Del}, &
    \mathcal{K} &= \frac{\nu}{\alpha\Lambda}(Y^{-\alpha/\Del}-1),
\end{align}
with the universal critical
exponents\cite{fisher1983,pelissetto2002,anisimov2000inbook,behnejad2010,sengers2009}
\begin{align}
    \nu &= 0.63, &\Del &= 0.5,\\
    \gamma &= 1.24, &\alpha &= 0.11.
\end{align}
For the crossover function $Y(\kappa)$, we adopt the
expression\cite{holtentwostatesupplement}
\begin{equation}\label{eq:crossoverfunction}
    Y(\kappa) = \left(1+\frac{\Lambda^2}{\kappa^2}\right)^{-\Del/(2\nu)}.
\end{equation}
Here the parameter $\kappa$ is the effective distance from the critical point, which is
inversely proportional to the correlation length of the order-parameter fluctuations.
When $\kappa \rightarrow 0$, at the critical point, $\Lambda^2/\kappa^2 \rightarrow
\infty$, and the crossover function represents the asymptotic scaling regime. When
$\Lambda^2/\kappa^2 \ll 1$, far away from the critical point, $Y(\kappa) \rightarrow 1$
and the fluctuations are negligible (mean-field regime).

The  Gibbs energy of mixing $\A$ becomes renormalized as
\begin{align}
    \frac{\Ax}{\Th} =~&2\biggl[\xx\ln \xx + (1-\xx)\ln(1-\xx) - \frac{\Wx}{4}(1-2\xx)^2\biggr]\notag\\
            &-\frac{1}{2}c_t^2 h_2^2\cal{K}.
\end{align}
The parameter $\kappa$ is calculated by iteration from the implicit
relation\cite{edison1998,povodyrev1999,vanthof2001,holtentwostatesupplement}
\begin{equation}\label{eq:kappa}
    \kappa^2 = c_t Y^{\nu/(2\Del)}\left[\frac{1}{2\xx(1-\xx)} - \Wx\right].
\end{equation}
The condition for chemical reaction equilibrium yields the condition for the equilibrium
concentration $x=x\ts{e}$,
\begin{align}\label{eq:xecr}
    \ln K - \dxxdx \left[\ln\left(\frac{\xx}{1-\xx}\right) + \Wx(1-2\xx)\right]\notag\\
    + \left(\xx-\frac{1}{2}\right)^2 \dWxdx
    - \frac{1}{2}\pypxd{k}{x}{W} = 0.
\end{align}
This equation is used to find the value of the low-density fraction, when the influence
of critical fluctuations is taken into account. We note that the equations for the
physical properties, given by Eqs.~(\ref{eq:Vh})--(\ref{eq:Cph}), remain valid in the
crossover regime, provided that the scaling densities and susceptibilities are calculated
with the crossover procedure.

\section{Description of the ST2 model with the equation of state}

\subsection{Two versions of ST2}\label{sec:ST2versions}
\begin{figure}
\includegraphics{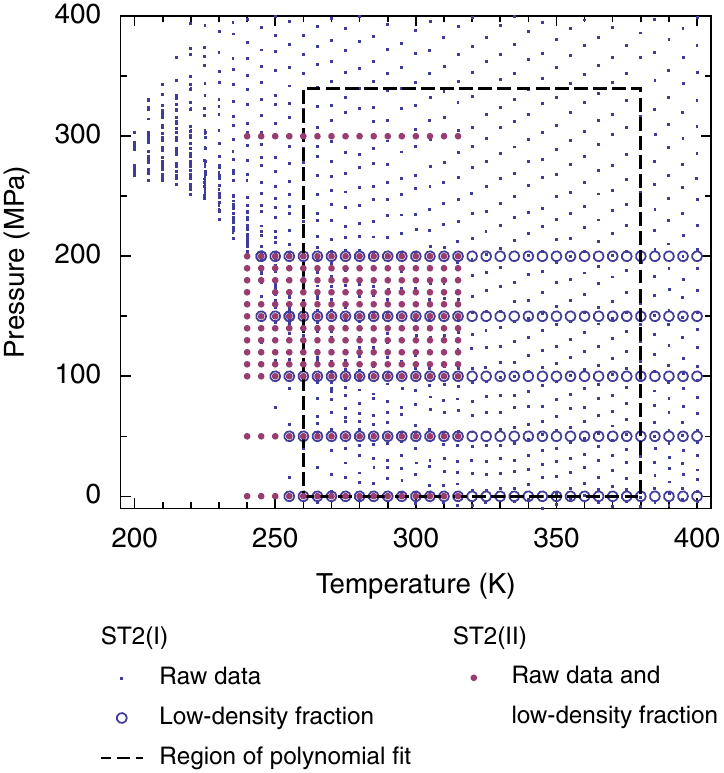}
\caption{\label{fig:phasediagramalldata}Location of the property data of the
\poolemodel and \palmermodel models. The raw data obtained for both models are values of energy and density
at each temperature and pressure point. For the \palmermodel model, the low-density fractions are also raw data.
For the \poolemodel data, polynomials are fitted
to the raw data for energy and density in the dashed region, to obtain values for the
density on isobars and values for the response functions. Values for the low-density
fraction on isobars are obtained by linear interpolation of the raw data.
}
\end{figure}

In this work, we consider two versions of the ST2 model of water.\cite{stillinger1974}
One version, which we refer to as \poolemodel, has been investigated in
Refs.~\onlinecite{poole2005,cuthbertsonpoole2011} with the reaction field method to
approximate electrostatic interactions. Data from the \poolemodel model, previously
published in Ref.~\onlinecite{poole2005}, were obtained from molecular dynamics
simulations of 1728 ST2 water molecules, at constant temperature and volume. Complete
details of the simulation procedure are as described in Ref.~\onlinecite{poole2005}. Raw
data for this model consist of pressure and energy as a function of volume and
temperature.  As a result, data on isobars must be obtained by interpolation.  To
facilitate this, we fitted to the $P(V,T)$ and $E(V,T)$ data a bivariate polynomial of
the form
\begin{equation}
    \sum_{i=0}^{7} \sum_{j=0}^{6} a_{ij} c_i(V) c_j(T),
\end{equation}
where $c_i(V)$ is the Chebyshev polynomial of the first kind of degree $i$ as a function
of $V$, $c_j(T)$ is similarly defined, and $a_{ij}$ are the parameters of the fit.  The
fits also allow us to estimate response functions via differentiation of the fitted
functions. The fit is valid only in a reduced temperature and pressure range, as shown in
Fig.~\ref{fig:phasediagramalldata}. In addition, the fraction of molecules in a
low-density state was estimated from the distance $r_5$ from the oxygen atom of a
molecule to its fifth-nearest neighbor.\cite{cuthbertsonpoole2011} Molecules were
assigned to the low-density state when $r_5 > 0.35$~nm, and to the high-density state
otherwise. The raw data for the fraction of molecules in the low-density state were
obtained as a function of volume and temperature, and were converted to data on isobars
by a linear interpolation between data points. The location of the interpolated points is
shown in Fig.~\ref{fig:phasediagramalldata}.

The second version of the ST2 model has been investigated in
Refs.~\onlinecite{liu2009,liu2012,palmer2013}, and will be referred to as \palmermodel.
Data from the \palmermodel model were calculated from Monte Carlo simulations of 400 ST2
water molecules at constant pressure and temperature, with the Ewald treatment of
electrostatics with vacuum boundary conditions. The raw data consist of density and
energy as a function of temperature and pressure, at locations shown in
Fig.~\ref{fig:phasediagramalldata}. In addition, we computed the low-density fraction at
each point, using the same criterion as for the \poolemodel model.

The approximations used in the formulation of the two-state thermodynamics, in particular
\eqref{eq:lnK}, make our equation of state less accurate away from the critical point.
This is why, to reduce the required number of background terms in \eqref{eq:background},
our equation of state was fitted to data in the reduced pressure ranges of 100~MPa to
250~MPa (\poolemodel) and 100~MPa to 300~MPa (\palmermodel). In the case of the
\poolemodel data, the temperature range was also reduced to the range of 240~K to 322~K.
For \poolemodel, the equation of state was fitted to the volumes and response functions
that were calculated from the polynomial fit, while for \palmermodel, the equation was
directly fitted to density and energy values obtained from simulations. The total number
of the adjustable background coefficients is thus restricted to seven (see tables
\ref{tab:PooleMF}--\ref{tab:PalmerCR} in Appendix~\ref{app:tables}). The parameter
$c_{00}$ defines the zero point of energy. For the current ST2 simulations, the zero
point of energy corresponds to moving all the molecules in the system infinitely far
apart from one another.

\subsection{Source of nonideality: energy-driven versus entropy-driven}
\begin{figure}
\includegraphics{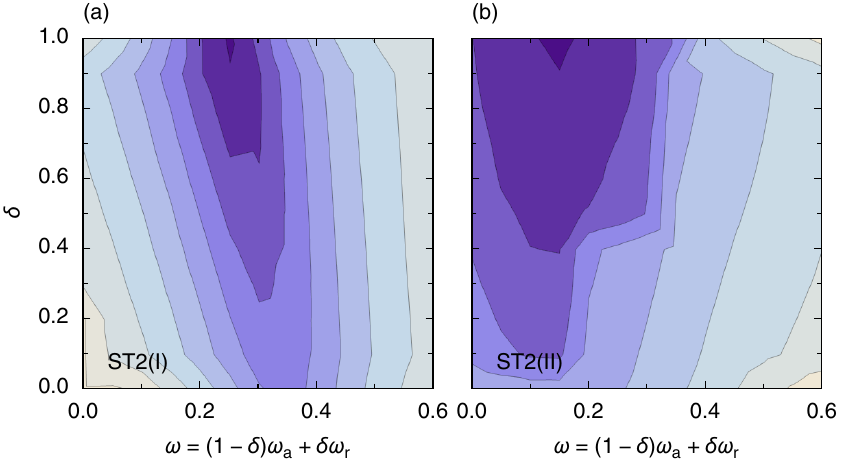}
\caption{\label{fig:contoursomegadelta}Contour lines of the sum of squared deviations $\chi^2$ of the
data from the equation of state as a function of the effective interaction parameter $\omega$
given by \eqref{eq:omega} and $\delta$, the switching parameter between entropy-driven and energy-driven
nonidealities, for the \poolemodel (a) and \palmermodel (b) models. Darker colors represent a better fit, where
in (a), the darkest color represents $\chi^2\leq2.5$, with subsequent contours corresponding to
increases in $\chi^2$ by a factor of 1.6. In (b), the darkest color represents $\chi^2\leq1.3$, and
subsequent contours correspond to increases in $\chi^2$ by a factor of 1.3.
For this analysis, the mean-field version of the equation of state is used,
and the critical point location is fixed as 252~K, 165~MPa for \poolemodel and
247~K, 155~MPa for \palmermodel.}
\end{figure}

Figure~\ref{fig:contoursomegadelta} shows the quality of the description of the ST2 data
of versions I and II, with the nonideality factor $W$ given by \eqref{eq:W}, as a
function of $\delta$ and the effective coefficient $\omega$,
\begin{equation}  \label{eq:omega}
\omega \equiv (1-\delta)\omega\ts{a} + \delta\omega\ts{r}.
\end{equation}
For any value of $\delta$, a pressure dependence of $W$ appears to be necessary, with an
optimum $\omega$ between 0.25 and 0.35. However, an energy-driven nonideality (with
$\delta=1$) gives a much better result than entropy-driven nonideality (with $\delta=0$).
Therefore, in subsequent sections, we adopt $\delta = 1$.

\subsection{Location of the liquid--liquid critical point}

\begin{figure}
\includegraphics{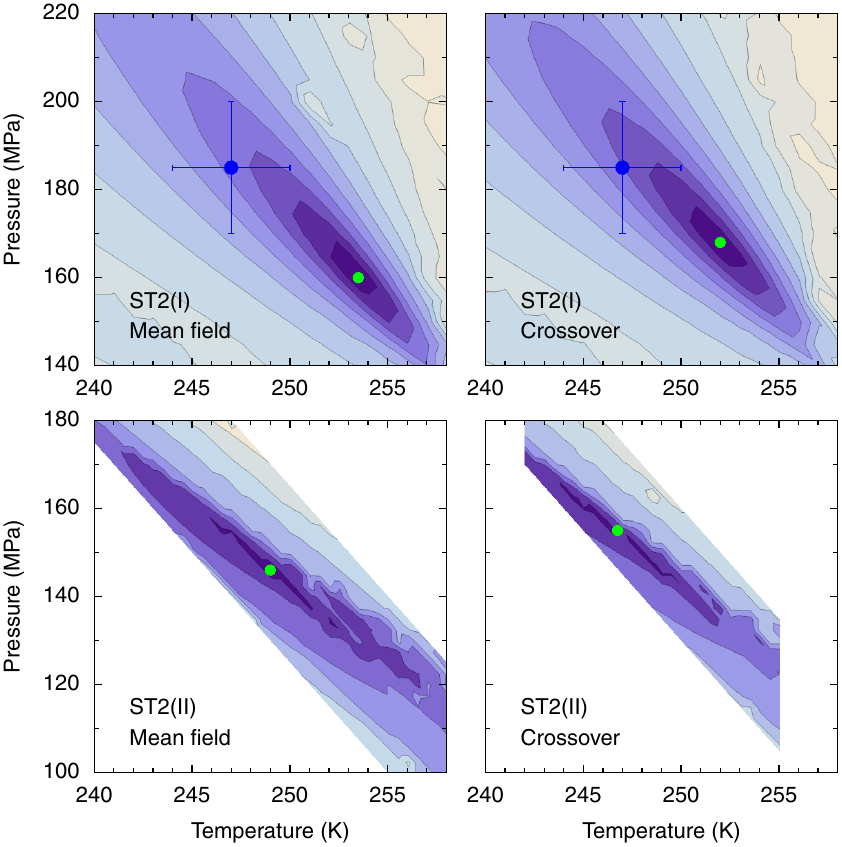}
\caption{\label{fig:criticalpointcontours}Location of the critical point for the mean-field
equation (left) and the crossover equation (right),
for the \poolemodel (top) and \palmermodel (bottom) models.
Contours indicate the sum of squared deviations $\chi^2$ of the fit from the data,
and darker colors represent lower deviations.
For \poolemodel, the darkest color represents $\chi^2\leq4$, with subsequent contours
corresponding to increases in $\chi^2$ by a factor of 1.6.
For \palmermodel, the darkest
color represents $\chi^2\leq0.4$, with subsequent contours corresponding to increases in
$\chi^2$ by a factor of 1.6.
The green dot shows the optimum critical point location.
The blue dot with error bars represents the estimate
by Cuthbertson and Poole\cite{cuthbertsonpoole2011}
of the location of the critical point.}
\end{figure}

\noindent Two approximations of the equation of state were considered: the mean-field
equation described in \secref{sec:meanfield}, and a renormalized equation of state that
takes into account critical fluctuations and exhibits critical scaling behavior close to
the critical point (\secref{sec:crossover}). The latter equation of state crosses over to
mean-field behavior away from the critical point. Figure~\ref{fig:criticalpointcontours}
shows how the goodness of the fit of both equations depends on the location of the
critical point, for both the \poolemodel and \palmermodel model. For both ST2 models, the
optimum critical point location for the crossover equation is at lower temperature and
higher pressure than that of the mean-field equation. For the \poolemodel model, the
location of the critical point predicted by the crossover equation is close to the
estimate of Cuthbertson and Poole,\cite{cuthbertsonpoole2011} although outside of the
reported error bars. Taking into account critical fluctuations causes a shift of the
critical point in the direction of the two-phase
region.\cite{anisimov1992,edison1998,povodyrev1999} The shift in pressure is about 8~MPa,
which is 7\% of $\rhoc R\Tc$.

\begin{figure}
\includegraphics{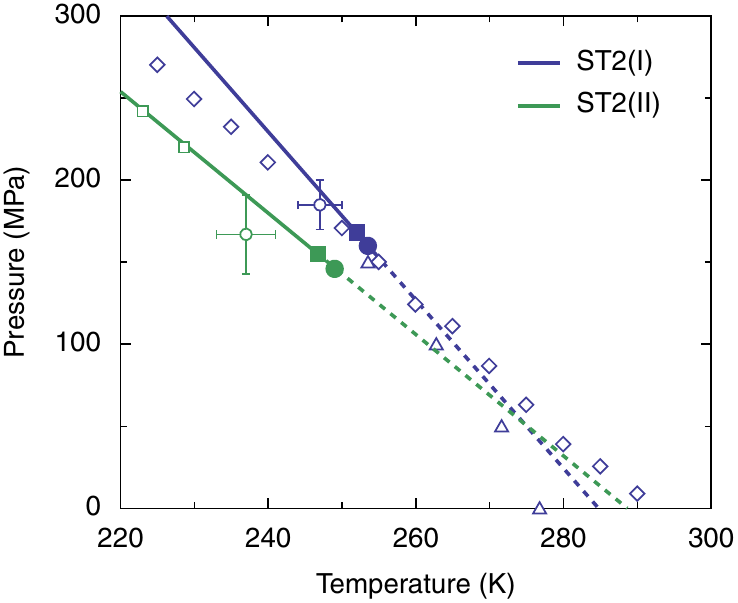}
\caption{\label{fig:phasediagrammodels}Comparison of critical point locations and
liquid--liquid transition lines for the \poolemodel and \palmermodel models. Open circles
(with error bars) are critical point locations reported in Refs~\onlinecite{cuthbertsonpoole2011,liu2009}.
Closed circles and squares: fitted location of critical point according
to the mean-field or crossover equation, respectively. The lines
are the liquid--liquid transition lines (solid) and Widom lines (dashed).
Open diamonds are points with a low-density fraction equal to
1/2 (\poolemodel).\cite{cuthbertsonpoole2011}
Triangles: inflection points in the low-density fraction versus temperature.
Open squares: estimated locations of phase transition\cite{liu2012} (\palmermodel)
}
\end{figure}

The Gibbs energy of mixing $\A$, given by \eqref{eq:deltaG}, is symmetric with respect to
the order parameter $\phi_1 = 2x-1$. This means that in our equation of state the Widom
line corresponds to the low-density fraction $x=1/2$ and to the inflection point in the
fraction versus temperature. The low-density fraction was also found to be be $1/2$ at
the Widom line in the simulated TIP4P/2005 model.\cite{wikfeldt2011} However, in our
equation of state this is a simplification, valid only close to the critical point.
\ffigref{fig:phasediagrammodels} shows that for \poolemodel, the locations of $x=1/2$ and
the inflection points become different upon departure from the critical point, with our
prediction being a compromise between these two. To improve this feature, one needs to
introduce some asymmetry with respect to the low-density fraction $x$ in the Gibbs energy
of mixing [\eqref{eq:deltaG}], and go beyond the linear approximation in the expression
for $\ln K$ [\eqref{eq:lnK}].

For the \palmermodel model, the shift in critical pressure between the mean-field and
crossover critical point locations is about 9~MPa (Fig.~\ref{fig:criticalpointcontours}).
The optimum critical temperatures that we find (249.0~K for the mean-field equation and
246.8~K for the crossover equation) do not agree with the 2009 estimate of Liu
\ea\cite{liu2009} of $237\pm4$~K shown in Fig.~\ref{fig:phasediagrammodels}. However, the
liquid--liquid transition line that we find does agree with the points at 223~K and
228.6~K that Liu \ea\cite{liu2012} estimated in 2012.

\subsection{Description of thermodynamic properties}
\begin{figure*}
\includegraphics{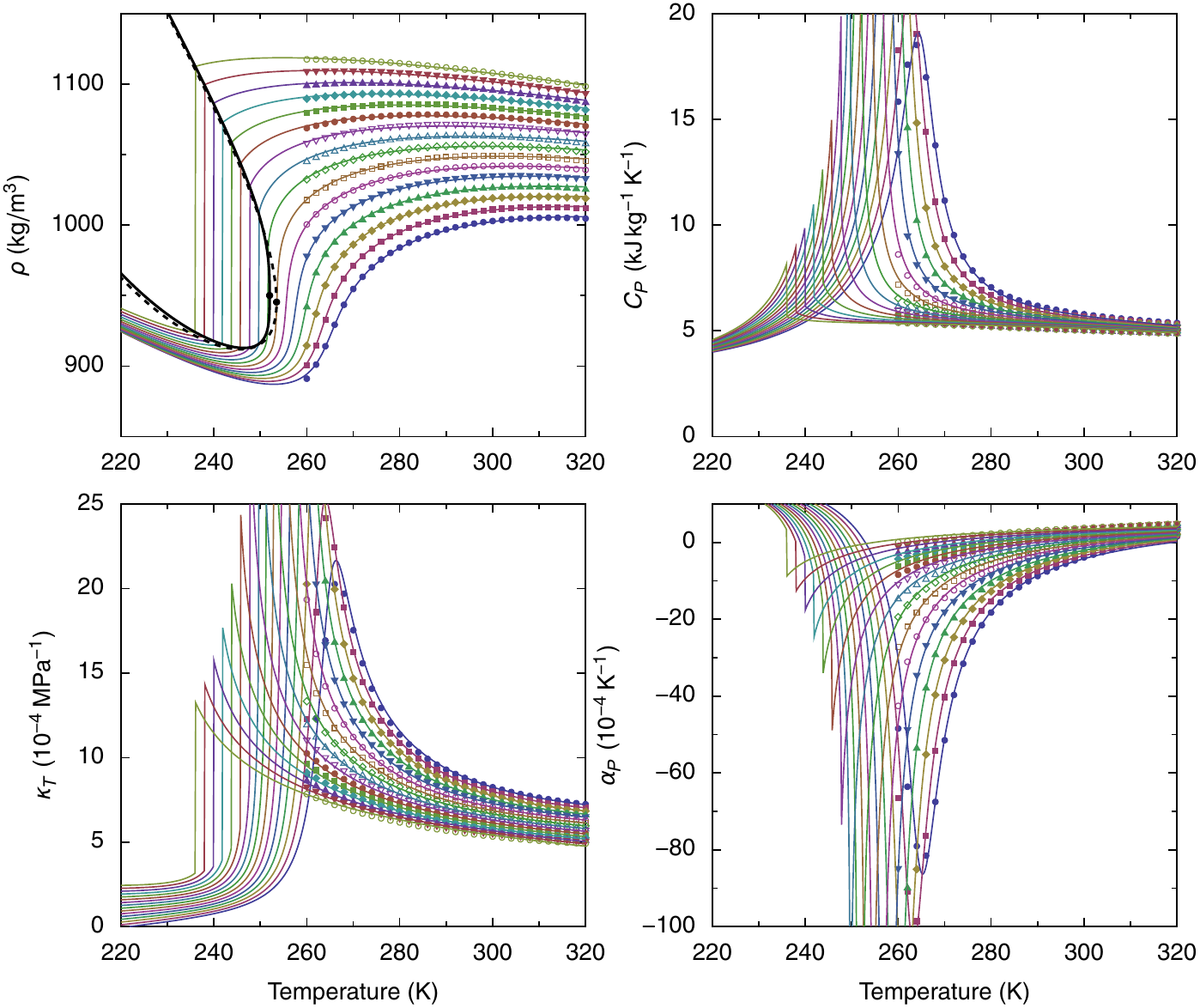}
\caption{Density $\protect\rho$, isobaric heat capacity $C_P$, isothermal compressibility
$\protect\kappa_T$, and expansivity $\protect\alpha_P$ along isobars, predicted by the
crossover equation of state. The data points are obtained from polynomial fits to raw
data for the volume and energy for the \poolemodel model. In the density graph, the black
curves indicate the phase coexistence densities (dashed: mean-field equation, solid:
crossover equation), and the black dots represent the mean-field and crossover locations
for the critical point. Isobar pressures are 100~MPa to 250~MPa in steps of 10~MPa.}
\label{fig:pooleproperties26}
\end{figure*}

\begin{figure}
\includegraphics{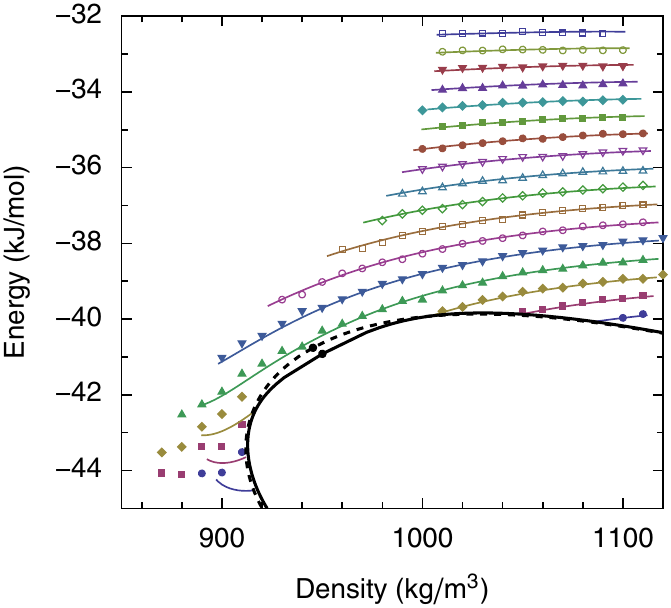}
\caption{\label{fig:Pooleenergyvsdensity}Isotherms of energy $E$ versus density $\rho$, for
the \poolemodel model (points) and predictions by the crossover equation of state.
The black curves indicate phase coexistence (dashed: mean-field equation,
solid: crossover equation),
and the black dots represent the mean-field and crossover locations
for the critical point. From bottom to top, the isotherm temperatures are 240~K to 320~K in steps of 5~K.}
\end{figure}

Figure~\ref{fig:pooleproperties26} shows thermodynamic-property data from the polynomial
fit of the \poolemodel model, together with predictions from the equation of state. The
predictions match the data well. In the region where \poolemodel data are available, the
fits of the mean-field and crossover equations are almost indistinguishable. This is why
the curves for the properties are shown only for the crossover equation, except for the
liquid--liquid coexistence curve, for which the shapes are different in the close
vicinity of the critical point. In Fig.~\ref{fig:Pooleenergyvsdensity}, the raw
\poolemodel data are shown in an energy--density plot. The molar energy $E$ is calculated
as
\begin{equation}
    E = G + T S - P V.
\end{equation}
The equation of state agrees well with the data, except for three isotherms at low
density and low temperatures.

\begin{figure*}
\includegraphics{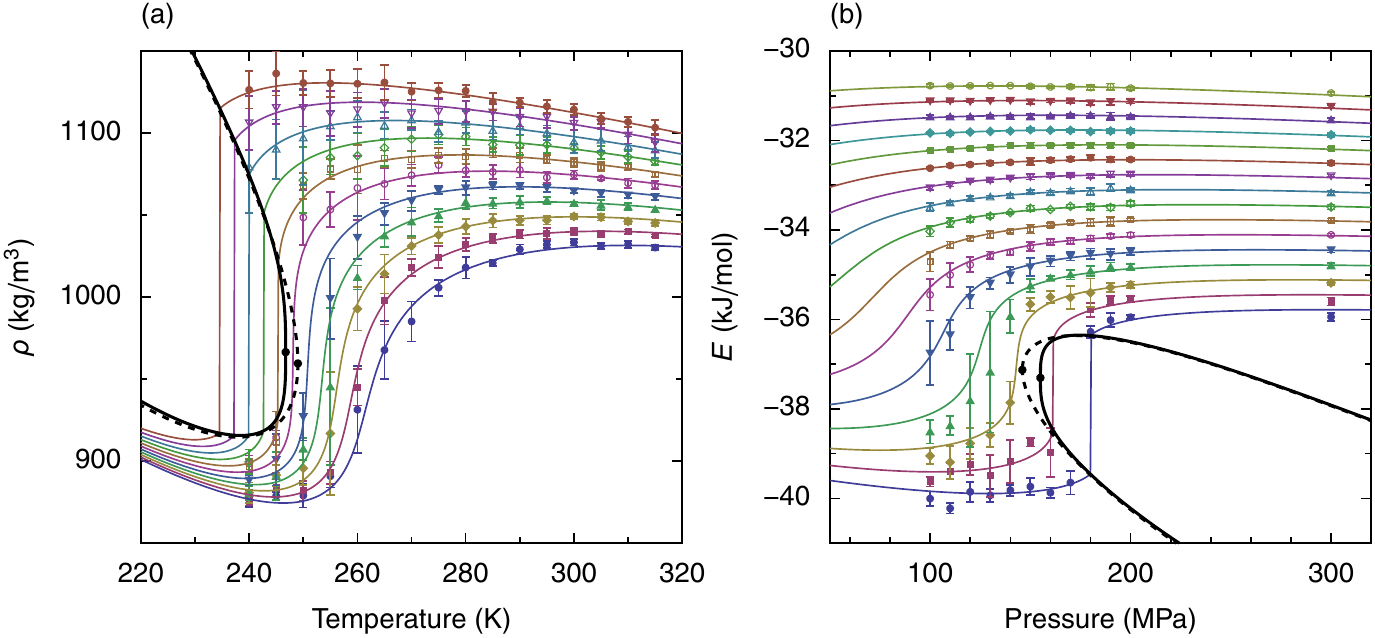}
\caption{\label{fig:Palmerdensityandenergy}Density $\protect\rho$ (a) and molar energy
$E$ (b) obtained for the \palmermodel model (points) and predicted by the crossover
equation of state. The black curves indicate two-phase coexistence (dashed: mean-field
equation, solid: crossover equation), and the black dots represent the critical point. In
(a) the isobar pressures are 100~MPa to 200~MPa in steps of 10~MPa. In (b), the isotherm
temperatures are 240~K to 315~K in steps of 5~K.}
\end{figure*}

\begin{figure}
\includegraphics{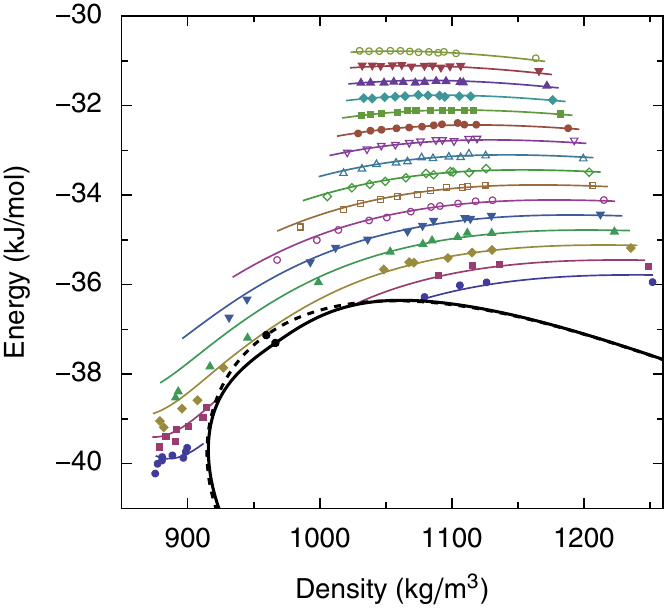}
\caption{\label{fig:Palmerenergyvsdensity}Isotherms of energy $E$ versus density $\rho$, for
the \palmermodel model (points) and predictions by the crossover two-state equation.
The black curves indicate phase coexistence (dashed: mean-field equation,
solid: crossover equation),
and the black dots represent the critical point.
From bottom to top, the isotherm temperatures are 240~K to 315~K in steps of 5~K.}
\end{figure}

For the \palmermodel model, the equation of state was fitted to the raw data. The results
are shown in Figs.~\ref{fig:Palmerdensityandenergy} and \ref{fig:Palmerenergyvsdensity}.
It is seen from both figures that the quality of the description of the density and
energy is even better than for the \poolemodel model, in particular for low density and
low temperature.

The numerical values of the parameters of the equation of state are given in tables
\ref{tab:PooleMF}--\ref{tab:PalmerCR}.

\subsection{Low-density fraction}
\begin{figure}
\includegraphics{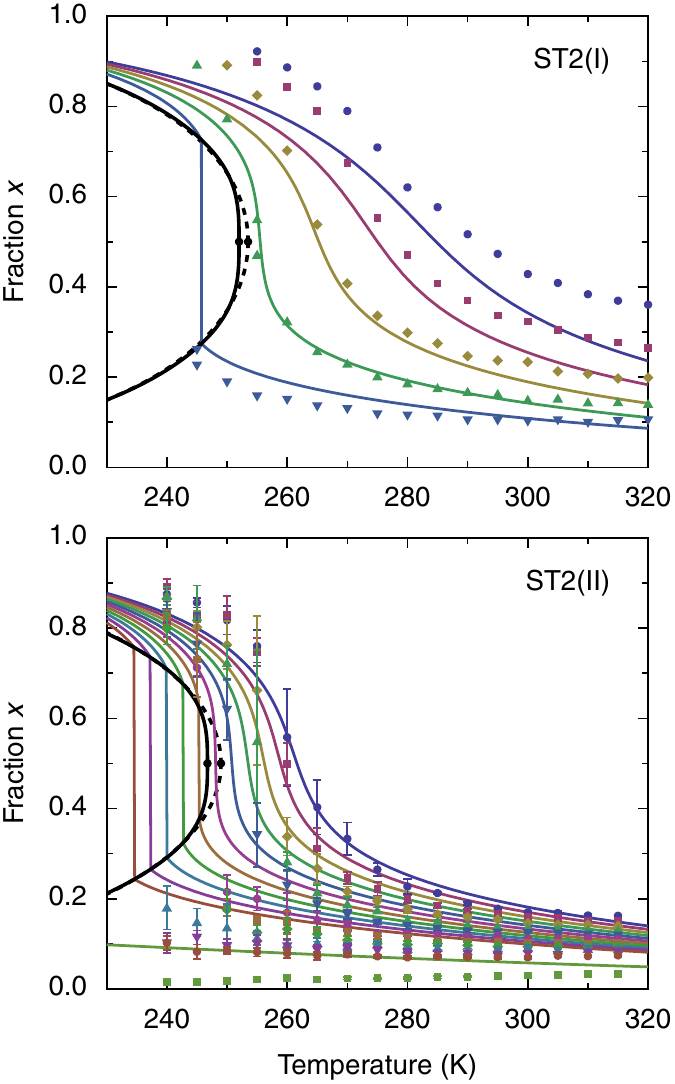}
\caption{\label{fig:fraction}
Low-density fraction from the \poolemodel and \palmermodel models
(points) and the predictions
from the crossover equation of state. The black curves indicate
the phase coexistence fractions  (dashed: mean-field equation,
solid: crossover equation).
For \poolemodel, the isobar pressures are 0~MPa to 200~MPa in steps of 50~MPa.
For \palmermodel, the isobar pressures are 100~MPa to 200~MPa in steps of 10~MPa,
followed by 300~MPa.
}
\end{figure}

The low-density fractions $x$ predicted (not fitted) by the equation of state are shown
in Fig.~\ref{fig:fraction}. The fractions predicted by the mean-field equation and the
crossover equation are indistinguishable, except for the fraction at the liquid--liquid
coexistence in the close vicinity to the critical point. For both ST2 models, the
predictions do qualitatively represent the low-density fraction. For \poolemodel, where
data are available close to the critical point, the predictions are quantitatively good.

\section{Discussion and conclusion}
\begin{figure}
\includegraphics{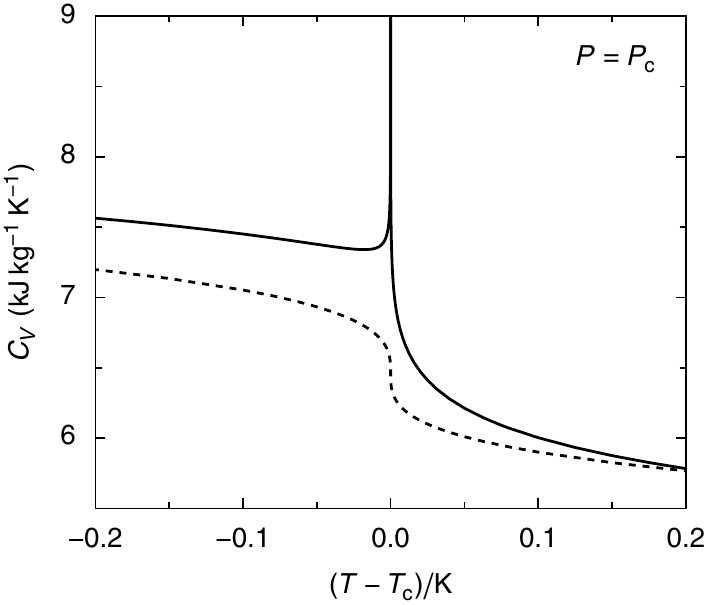}
\caption{\label{fig:palmerCV}
Heat capacity at constant volume $C_V$ along the critical isobar, $P=\Pc$,
predicted for the \palmermodel model for the mean-field equation (dashed)
and crossover equation (solid).}
\end{figure}

Our analysis of the two versions of the ST2 model could not detect significant deviations
from the mean-field approximation of the equation of state. To unambiguously detect the
effects of the critical fluctuations, we would need more data close to the critical
point. With the data currently available, the location of the critical point for both
versions of the ST2 model is still not certain enough. Secondly, for the properties
analyzed the effects of fluctuations are not very pronounced. There is one property whose
anomaly solely originates from the fluctuation effects. This is the heat capacity at
constant volume $C_V$ which weakly diverges at the critical point in scaling theory and
remains finite in the mean-field approximation.\cite{anisimov2000inbook,behnejad2010} We
have calculated $C_V$ as a function of temperature along the critical isobar. The result
is shown in \figref{fig:palmerCV} for the \palmermodel model. In the crossover equation,
the peak of $C_V$ is very narrow (within 0.1~K). This feature makes such a peak
undetectable within the accuracy of the simulations, which have temperature increments of
5~K. For real supercooled water, it was also impossible to distinguish between the
mean-field approximation and effects of the critical fluctuations, because of the lack of
experimental data in the critical region. A parametric equation of state which obeys the
asymptotic scaling laws was applied to describe the thermodynamic properties of
supercooled water in Refs.~\onlinecite{fuentevilla2006,bertrand2011,holtenSCW}. The
crossover to mean-field behavior away from the critical point was absorbed in the
adjustable backgrounds. The quality of the description was similar to that of the
two-state equation.\cite{holtentwostate}

The order parameter in our equation of state, $\phi_1$, is introduced phenomenologically
through the low-density fraction. We assume that the order parameter thermodynamically
belongs to the same universality class as the order parameter near the vapor--liquid
transition (associated with the density). However, because the low-density fraction
relaxes through a reorientation of molecular bonds, dynamically, the order parameter may
belong to the nonconserved-order-parameter universality
class,\cite{tanaka1999,biddle2013} even if this order parameter is coupled with conserved
properties, density and energy, through \eqsref{eq:Vh}{eq:Sh}.

The results presented in previous sections show that the two-state physics reasonably
describes the properties of both versions of the ST2 model of water. It is interesting to
compare the application of this approach to two alternative models of water, ST2 and mW.
While the nonideality in mixing of the two states in the mW model is not strong enough to
cause a metastable liquid--liquid separation, the ST2 model does show the metastable
liquid--liquid transition, terminated by a critical point. Phenomenologically, the reason
for the different behavior of these two models may be associated with the source of the
nonideality. As shown in Ref.~\onlinecite{holtenmW}, in the mW model the nonideality is
purely entropy-driven (athermal mixing of two structures), while in the ST2 model the
nonideality is mainly energy-driven. As discussed in Ref.~\onlinecite{holtenmW}, the
physical origin of the excess entropy of mixing in the mW model can be attributed to the
clustering of the molecules involved in the low-density structure. The clustering
significantly suppresses the entropy of mixing. A similar effect is well known for
mixtures of high-molecular-weight polymers, where the number of statistical
configurations is restricted by polymerization.\cite{florybook} However, this effect
alone cannot cause phase separation without some contribution from the enthalpy of
mixing, which is almost nonexistent in the mW model.\cite{holtenmW}

We remain agnostic with respect to the most intriguing question: whether real liquid
water exhibits a metastable liquid--liquid transition or not. We do not, in other words,
address the question of whether real liquid water is closer to mW or to ST2. It was shown
in Ref.~\onlinecite{holtentwostate} that the anomalies of supercooled water can be
described by the nearly athermal, entropy-driven nonideality of mixing of the two states,
similar to mW. However, is the magnitude of the nonideality sufficient to cause the
metastable liquid--liquid separation? To answer this question, significantly more
accurate experiments on the response functions, as well as independent measurements of
the low-density fraction as a function of temperature and pressure in real water are
desirable.

\begin{acknowledgments}
Three of us gratefully acknowledge the support of the National Science Foundation (V.H.
and M.A.A.: Grant No. CHE-1012052; P.G.D.: Grant No. CHE-1213343). P.H.P. is supported by
the Natural Sciences and Engineering Research Council of Canada and ACEnet. M.A.A. thanks
Valeria Molinero for stimulating discussions and Limei Xu for suggesting additional
references.
\end{acknowledgments}

\appendix
\section{Derivation of Eqs. (\ref{eq:Vh})--(\ref{eq:Cph})} \label{app:derivation}
Critical phenomena can be characterized by the scaling fields $h_1$ and $h_2$, which are
given for our equation of state by \eqsref{eq:h1}{eq:h2},
\begin{align}
    h_1 &= \ln K,\\
    h_2 &= 2 - W,
\end{align}
and a dependent scaling field $h_3(h_1,h_2)$, which is the critical part of the
field-dependent thermodynamic potential. For our equation of state, we adopt the
form\cite{holtentwostatesupplement}
\begin{equation}
    h_3 = -\frac{2(\Gh - \GA)}{\Th} - \ln K + \frac{W}{2}.\label{eq:h3}
\end{equation}
The scaling fields are connected to the scaling densities $\phi_1$ and $\phi_2$ by the
relation
\begin{equation}
    \di h_3 = \phi_1 \di h_1 + \phi_2 \di h_2, \label{eq:h3dif}
\end{equation}
so that
\begin{equation}
    \phi_1  = \pypxd{h_3}{h_1}{h_2},\qquad \phi_2  = \pypxd{h_3}{h_2}{h_1}.
\end{equation}
All thermodynamic properties can be written as derivatives of the dimensionless Gibbs
energy $\Gh$. From \eqref{eq:h3} it follows that
\begin{equation}
    \Gh = -\frac{\Th}{2}\left(h_3 + \ln K - \frac{W}{2} \right) + \GA.
\end{equation}
The dimensionless volume $\Vh$ can then be written as
\begin{equation}\label{eq:Vhstep1}
    \Vh = \pypxd{\Gh}{\Ph}{\Th}
        = -\frac{\Th}{2}\left[\pypxd{h_3}{\Ph}{\Th} + a\lambda - \frac{\Wp}{2} \right] + \GA_{\Ph}.
\end{equation}
The pressure derivative of $h_3$ is found from \eqref{eq:h3dif},
\begin{align}
    \pypxd{h_3}{\Ph}{\Th} &= \phi_1 \pypxd{h_1}{\Ph}{\Th} + \phi_2 \pypxd{h_2}{\Ph}{\Th}\notag\\
                          &= \phi_1 a\lambda - \phi_2 \Wp. \label{eq:dh3dP}
\end{align}
When we substitute this result in \eqref{eq:Vhstep1}, we obtain \eqref{eq:Vh}. The
dimensionless entropy $\Sh$ is related to the Gibbs energy as
\begin{align}
    \Sh &= -\pypxd{\Gh}{\Th}{\Ph} \notag\\
        &= \frac{\Th}{2}\left[\pypxd{h_3}{\Th}{\Ph} + \lambda - \frac{\Wt}{2}\right]
            + \frac{1}{2}\left(h_3 + \ln K - \frac{W}{2} \right) - \GA_{\Th}\notag\\
        &= \frac{\Th}{2}\left[\phi_1 \lambda - \phi_2 \Wt + \lambda - \frac{\Wt}{2}\right]
            -\frac{\Gh-\GA}{\Th} - \GA_{\Th},
\end{align}
where the temperature derivative of $h_3$ was obtained analogously to \eqref{eq:dh3dP}.
This result for $\Sh$ leads to \eqref{eq:Sh}.

The dimensionless response functions are expressed in terms of the scaling
susceptibilities $\chi_1$, $\chi_{12}$, and $\chi_2$, which are connected to the scaling
densities by the relations
\begin{align}
    \di \phi_1 &= \chi_1    \di h_1 + \chi_{12} \di h_2,\label{eq:phi1dif}\\
    \di \phi_2 &= \chi_{12} \di h_1 + \chi_2    \di h_2.\label{eq:phi2dif}
\end{align}
The dimensionless compressibility $\kap$ is related to $\Vh$ as
\begin{equation}\label{eq:kapstep1}
    \kap\Vh = -\pypxd{\Vh}{\Ph}{\Th}
            = \frac{\Th}{2}\left[a\lambda\pypxd{\phi_1}{\Ph}{\Th}
                - \Wp\pypxd{\phi_2}{\Ph}{\Th}\right] - \GA_{\Ph\Ph},
\end{equation}
under the assumption that $\Wpp = 0$. The pressure derivative of $\phi_1$ is found from
\eqref{eq:phi1dif},
\begin{align}
    \pypxd{\phi_1}{\Ph}{\Th} &= \chi_1 \pypxd{h_1}{\Ph}{\Th} + \chi_2 \pypxd{h_2}{\Ph}{\Th}\notag\\
                             &= \chi_1 a\lambda + \chi_{12}\Wp,
\end{align}
and the pressure derivative of $\phi_2$ is found analogously from \eqref{eq:phi2dif}.

The dimensionless expansivity $\alp$ and heat capacity $\Cph$ are found from the
relations
\begin{equation}
    \alp\Vh = \pypxd{\Vh}{\Th}{\Ph}, \qquad \frac{\Cph}{\Th} = \pypxd{\Sh}{\Th}{\Ph},
\end{equation}
which are evaluated in the same manner as \eqref{eq:kapstep1}.

\section{Tables}\label{app:tables}
For all fits, the parameter $c_{01}$ is given by
\begin{equation}
    c_{01} = 1 + a \lambda /2 - \omega\ts{r}/4,
\end{equation}
to ensure that $\Vh = 1$ at the critical point. The parameter $c_{00}$ defines the zero
point of energy, as described in \secref{sec:ST2versions}. The parameter $c_{10}$ defines
the zero point of entropy and was taken zero. The molecular cutoff parameter in the
crossover procedure is optimized by trial at the value $\Lambda = 0.3$ for both versions
of the ST2 model.

\begin{table}[h]
\caption{Parameters for the mean-field equation fitted to \poolemodel\label{tab:PooleMF}}
\begin{ruledtabular}
\begin{tabular}{lD{.}{.}{12}lD{.}{.}{12}}
Parameter & \multicolumn{1}{c}{Value} & Parameter & \multicolumn{1}{c}{Value}\\\hline
$\Tc$	&	253.5	~\text{K}	&	$c_{02}$	&	-7.275\;1\times 10^{-3}	\\
$\Pc$	&	160.0	~\text{MPa}	&	$c_{03}$	&	-2.330\;2\times 10^{-3}	\\
$\rhoc$	&	52.478	~\text{kmol m$^{-3}$}	&	$c_{11}$	&	1.950\;3\times 10^{-1}	\\
$\lambda$	&	-3.491\,5		&	$c_{12}$	&	-2.765\;1\times 10^{-2}	\\
$a$	&	0.085\,811		&	$c_{13}$	&	1.202\;0\times 10^{-2}	\\
$\omega\ts{r}$	&	0.231\,65		&	$c_{20}$	&	-5.830\;3	\\
$c_{00}$	&	-1.664\;6\times 10^{1}		&	$c_{30}$	&	2.115\;0	\\
\end{tabular}
\end{ruledtabular}
\end{table}

\begin{table}[h]
\caption{Parameters for the crossover equation fitted to \poolemodel\label{tab:PooleCR}}
\begin{ruledtabular}
\begin{tabular}{lD{.}{.}{12}lD{.}{.}{12}}
$\Tc$	&	252.0	~\text{K}	&	$c_{02}$	&	-8.350\;6\times 10^{-3}	\\
$\Pc$	&	168.0	~\text{MPa}	&	$c_{03}$	&	-2.006\;6\times 10^{-3}	\\
$\rhoc$	&	52.739	~\text{kmol m$^{-3}$}	&	$c_{11}$	&	1.855\;0\times 10^{-1}	\\
$\lambda$	&	-3.464\,3		&	$c_{12}$	&	-2.145\;7\times 10^{-2}	\\
$a$	&	0.085\,496		&	$c_{13}$	&	1.078\;2\times 10^{-2}	\\
$\omega\ts{r}$	&	0.238\,02		&	$c_{20}$	&	-5.856\;6	\\
$c_{00}$	&	-1.677\;7\times 10^{1}		&	$c_{30}$	&	2.092\;5	\\
\end{tabular}
\end{ruledtabular}
\end{table}

\begin{table}[h]
\caption{Parameters for the mean-field equation fitted to
\palmermodel\label{tab:PalmerMF}}
\begin{ruledtabular}
\begin{tabular}{lD{.}{.}{12}lD{.}{.}{12}}
Parameter & \multicolumn{1}{c}{Value} & Parameter & \multicolumn{1}{c}{Value}\\\hline
$\Tc$	&	249.0	~\text{K}	&	$c_{02}$	&	-8.524\;1\times 10^{-3}	\\
$\Pc$	&	146.0	~\text{MPa}	&	$c_{03}$	&	-1.323\;2\times 10^{-3}	\\
$\rhoc$	&	53.261	~\text{kmol m$^{-3}$}	&	$c_{11}$	&	2.405\;2\times 10^{-1}	\\
$\lambda$	&	-3.202\,1		&	$c_{12}$	&	-1.199\;9\times 10^{-2}	\\
$a$	&	0.119\,9		&	$c_{13}$	&	3.415\;1\times 10^{-3}	\\
$\omega\ts{r}$	&	0.171\,81		&	$c_{20}$	&	-4.400\;7	\\
$c_{00}$	&	-1.550\;9\times 10^{1}		&	$c_{30}$	&	1.768\;5	\\
\end{tabular}
\end{ruledtabular}
\end{table}

\begin{table}
\caption{Parameters for the crossover equation fitted to
\palmermodel\label{tab:PalmerCR}}
\begin{ruledtabular}
\begin{tabular}{lD{.}{.}{12}lD{.}{.}{12}}
Parameter & \multicolumn{1}{c}{Value} & Parameter & \multicolumn{1}{c}{Value}\\\hline
$\Tc$	&	246.75	~\text{K}	&	$c_{02}$	&	-9.471\;7\times 10^{-3}	\\
$\Pc$	&	155.0	~\text{MPa}	&	$c_{03}$	&	-1.099\;9\times 10^{-3}	\\
$\rhoc$	&	53.645	~\text{kmol m$^{-3}$}	&	$c_{11}$	&	2.340\;0\times 10^{-1}	\\
$\lambda$	&	-3.179\,3		&	$c_{12}$	&	-8.685\;3\times 10^{-3}	\\
$a$	&	0.120\,53		&	$c_{13}$	&	2.836\;9\times 10^{-3}	\\
$\omega\ts{r}$	&	0.171\,81		&	$c_{20}$	&	-4.408\;8	\\
$c_{00}$	&	-1.568\;5\times 10^{1}		&	$c_{30}$	&	1.714\;6	\\
\end{tabular}
\end{ruledtabular}
\end{table}

\bibliography{vincentnew,supercooled}

\begin{thebibliography}{76}%
\makeatletter
\providecommand \@ifxundefined [1]{%
 \@ifx{#1\undefined}
}%
\providecommand \@ifnum [1]{%
 \ifnum #1\expandafter \@firstoftwo
 \else \expandafter \@secondoftwo
 \fi
}%
\providecommand \@ifx [1]{%
 \ifx #1\expandafter \@firstoftwo
 \else \expandafter \@secondoftwo
 \fi
}%
\providecommand \natexlab [1]{#1}%
\providecommand \enquote  [1]{``#1''}%
\providecommand \bibnamefont  [1]{#1}%
\providecommand \bibfnamefont [1]{#1}%
\providecommand \citenamefont [1]{#1}%
\providecommand \href@noop [0]{\@secondoftwo}%
\providecommand \href [0]{\begingroup \@sanitize@url \@href}%
\providecommand \@href[1]{\@@startlink{#1}\@@href}%
\providecommand \@@href[1]{\endgroup#1\@@endlink}%
\providecommand \@sanitize@url [0]{\catcode `\\12\catcode `\$12\catcode
  `\&12\catcode `\#12\catcode `\^12\catcode `\_12\catcode `\%12\relax}%
\providecommand \@@startlink[1]{}%
\providecommand \@@endlink[0]{}%
\providecommand \url  [0]{\begingroup\@sanitize@url \@url }%
\providecommand \@url [1]{\endgroup\@href {#1}{\urlprefix }}%
\providecommand \urlprefix  [0]{URL }%
\providecommand \Eprint [0]{\href }%
\providecommand \doibase [0]{http://dx.doi.org/\ignorespaces}%
\providecommand \selectlanguage [0]{\@gobble}%
\providecommand \bibinfo  [0]{\@secondoftwo}%
\providecommand \bibfield  [0]{\@secondoftwo}%
\providecommand \translation [1]{[#1]}%
\providecommand \BibitemOpen [0]{}%
\providecommand \bibitemStop [0]{}%
\providecommand \bibitemNoStop [0]{.\EOS\space}%
\providecommand \EOS [0]{\spacefactor3000\relax}%
\providecommand \BibitemShut  [1]{\csname bibitem#1\endcsname}%
\let\auto@bib@innerbib\@empty
\bibitem [{\citenamefont {Angell}, \citenamefont {Shuppert},\ and\
  \citenamefont {Tucker}(1973)}]{angell1973}%
  \BibitemOpen
  \bibfield  {author} {\bibinfo {author} {\bibfnamefont {C.~A.}\ \bibnamefont
  {Angell}}, \bibinfo {author} {\bibfnamefont {J.}~\bibnamefont {Shuppert}}, \
  and\ \bibinfo {author} {\bibfnamefont {J.~C.}\ \bibnamefont {Tucker}},\
  }\href@noop {} {\bibfield  {journal} {\bibinfo  {journal} {J. Phys. Chem.}\
  }\textbf {\bibinfo {volume} {77}},\ \bibinfo {pages} {3092} (\bibinfo {year}
  {1973})}\BibitemShut {NoStop}%
\bibitem [{\citenamefont {Angell}, \citenamefont {Oguni},\ and\ \citenamefont
  {Sichina}(1982)}]{angell1982}%
  \BibitemOpen
  \bibfield  {author} {\bibinfo {author} {\bibfnamefont {C.~A.}\ \bibnamefont
  {Angell}}, \bibinfo {author} {\bibfnamefont {M.}~\bibnamefont {Oguni}}, \
  and\ \bibinfo {author} {\bibfnamefont {W.~J.}\ \bibnamefont {Sichina}},\
  }\href@noop {} {\bibfield  {journal} {\bibinfo  {journal} {J. Phys. Chem.}\
  }\textbf {\bibinfo {volume} {86}},\ \bibinfo {pages} {998} (\bibinfo {year}
  {1982})}\BibitemShut {NoStop}%
\bibitem [{\citenamefont {Tombari}, \citenamefont {Ferrari},\ and\
  \citenamefont {Salvetti}(1999)}]{tombari1999}%
  \BibitemOpen
  \bibfield  {author} {\bibinfo {author} {\bibfnamefont {E.}~\bibnamefont
  {Tombari}}, \bibinfo {author} {\bibfnamefont {C.}~\bibnamefont {Ferrari}}, \
  and\ \bibinfo {author} {\bibfnamefont {G.}~\bibnamefont {Salvetti}},\
  }\href@noop {} {\bibfield  {journal} {\bibinfo  {journal} {Chem. Phys.
  Lett.}\ }\textbf {\bibinfo {volume} {300}},\ \bibinfo {pages} {749} (\bibinfo
  {year} {1999})}\BibitemShut {NoStop}%
\bibitem [{\citenamefont {Archer}\ and\ \citenamefont {Carter}(2000)}]{arc00}%
  \BibitemOpen
  \bibfield  {author} {\bibinfo {author} {\bibfnamefont {D.~G.}\ \bibnamefont
  {Archer}}\ and\ \bibinfo {author} {\bibfnamefont {R.~W.}\ \bibnamefont
  {Carter}},\ }\href@noop {} {\bibfield  {journal} {\bibinfo  {journal}
  {J.~Phys. Chem.~B}\ }\textbf {\bibinfo {volume} {104}},\ \bibinfo {pages}
  {8563} (\bibinfo {year} {2000})}\BibitemShut {NoStop}%
\bibitem [{\citenamefont {Speedy}\ and\ \citenamefont
  {Angell}(1976)}]{speedy1976}%
  \BibitemOpen
  \bibfield  {author} {\bibinfo {author} {\bibfnamefont {R.~J.}\ \bibnamefont
  {Speedy}}\ and\ \bibinfo {author} {\bibfnamefont {C.~A.}\ \bibnamefont
  {Angell}},\ }\href@noop {} {\bibfield  {journal} {\bibinfo  {journal}
  {J.~Chem. Phys.}\ }\textbf {\bibinfo {volume} {65}},\ \bibinfo {pages} {851}
  (\bibinfo {year} {1976})}\BibitemShut {NoStop}%
\bibitem [{\citenamefont {Kanno}\ and\ \citenamefont
  {Angell}(1979)}]{kanno1979}%
  \BibitemOpen
  \bibfield  {author} {\bibinfo {author} {\bibfnamefont {H.}~\bibnamefont
  {Kanno}}\ and\ \bibinfo {author} {\bibfnamefont {C.~A.}\ \bibnamefont
  {Angell}},\ }\href@noop {} {\bibfield  {journal} {\bibinfo  {journal}
  {J.~Chem. Phys.}\ }\textbf {\bibinfo {volume} {70}},\ \bibinfo {pages} {4008}
  (\bibinfo {year} {1979})}\BibitemShut {NoStop}%
\bibitem [{\citenamefont {Mishima}(2010{\natexlab{a}})}]{mishima2010}%
  \BibitemOpen
  \bibfield  {author} {\bibinfo {author} {\bibfnamefont {O.}~\bibnamefont
  {Mishima}},\ }\href@noop {} {\bibfield  {journal} {\bibinfo  {journal}
  {J.~Chem. Phys.}\ }\textbf {\bibinfo {volume} {133}},\ \bibinfo {pages}
  {144503} (\bibinfo {year} {2010}{\natexlab{a}})}\BibitemShut {NoStop}%
\bibitem [{\citenamefont {Hare}\ and\ \citenamefont {Sorensen}(1987)}]{hare87}%
  \BibitemOpen
  \bibfield  {author} {\bibinfo {author} {\bibfnamefont {D.~E.}\ \bibnamefont
  {Hare}}\ and\ \bibinfo {author} {\bibfnamefont {C.~M.}\ \bibnamefont
  {Sorensen}},\ }\href@noop {} {\bibfield  {journal} {\bibinfo  {journal}
  {J.~Chem. Phys.}\ }\textbf {\bibinfo {volume} {87}},\ \bibinfo {pages} {4840}
  (\bibinfo {year} {1987})}\BibitemShut {NoStop}%
\bibitem [{\citenamefont {Kanno}\ and\ \citenamefont
  {Angell}(1980)}]{kanno1980}%
  \BibitemOpen
  \bibfield  {author} {\bibinfo {author} {\bibfnamefont {H.}~\bibnamefont
  {Kanno}}\ and\ \bibinfo {author} {\bibfnamefont {C.~A.}\ \bibnamefont
  {Angell}},\ }\href@noop {} {\bibfield  {journal} {\bibinfo  {journal}
  {J.~Chem. Phys.}\ }\textbf {\bibinfo {volume} {73}},\ \bibinfo {pages} {1940}
  (\bibinfo {year} {1980})}\BibitemShut {NoStop}%
\bibitem [{\citenamefont {Ter~Minassian}, \citenamefont {Pruzan},\ and\
  \citenamefont {Soulard}(1981)}]{terminassian1981}%
  \BibitemOpen
  \bibfield  {author} {\bibinfo {author} {\bibfnamefont {L.}~\bibnamefont
  {Ter~Minassian}}, \bibinfo {author} {\bibfnamefont {P.}~\bibnamefont
  {Pruzan}}, \ and\ \bibinfo {author} {\bibfnamefont {A.}~\bibnamefont
  {Soulard}},\ }\href@noop {} {\bibfield  {journal} {\bibinfo  {journal}
  {J.~Chem. Phys.}\ }\textbf {\bibinfo {volume} {75}},\ \bibinfo {pages} {3064}
  (\bibinfo {year} {1981})}\BibitemShut {NoStop}%
\bibitem [{\citenamefont {Poole}\ \emph {et~al.}(1992)\citenamefont {Poole},
  \citenamefont {Sciortino}, \citenamefont {Essmann},\ and\ \citenamefont
  {Stanley}}]{poole1992}%
  \BibitemOpen
  \bibfield  {author} {\bibinfo {author} {\bibfnamefont {P.~H.}\ \bibnamefont
  {Poole}}, \bibinfo {author} {\bibfnamefont {F.}~\bibnamefont {Sciortino}},
  \bibinfo {author} {\bibfnamefont {U.}~\bibnamefont {Essmann}}, \ and\
  \bibinfo {author} {\bibfnamefont {H.~E.}\ \bibnamefont {Stanley}},\
  }\href@noop {} {\bibfield  {journal} {\bibinfo  {journal} {Nature (London)}\
  }\textbf {\bibinfo {volume} {360}},\ \bibinfo {pages} {324} (\bibinfo {year}
  {1992})}\BibitemShut {NoStop}%
\bibitem [{\citenamefont {Mishima}\ and\ \citenamefont
  {Stanley}(1998{\natexlab{a}})}]{mishima1998review}%
  \BibitemOpen
  \bibfield  {author} {\bibinfo {author} {\bibfnamefont {O.}~\bibnamefont
  {Mishima}}\ and\ \bibinfo {author} {\bibfnamefont {H.~E.}\ \bibnamefont
  {Stanley}},\ }\href@noop {} {\bibfield  {journal} {\bibinfo  {journal}
  {Nature}\ }\textbf {\bibinfo {volume} {396}},\ \bibinfo {pages} {329}
  (\bibinfo {year} {1998}{\natexlab{a}})}\BibitemShut {NoStop}%
\bibitem [{\citenamefont {Stanley}\ \emph {et~al.}(2000)\citenamefont
  {Stanley}, \citenamefont {Buldyrev}, \citenamefont {Canpolat}, \citenamefont
  {Mishima}, \citenamefont {Sadr-Lahijany}, \citenamefont {Scala},\ and\
  \citenamefont {Starr}}]{stanley2000}%
  \BibitemOpen
  \bibfield  {author} {\bibinfo {author} {\bibfnamefont {H.~E.}\ \bibnamefont
  {Stanley}}, \bibinfo {author} {\bibfnamefont {S.~V.}\ \bibnamefont
  {Buldyrev}}, \bibinfo {author} {\bibfnamefont {M.}~\bibnamefont {Canpolat}},
  \bibinfo {author} {\bibfnamefont {O.}~\bibnamefont {Mishima}}, \bibinfo
  {author} {\bibfnamefont {M.~R.}\ \bibnamefont {Sadr-Lahijany}}, \bibinfo
  {author} {\bibfnamefont {A.}~\bibnamefont {Scala}}, \ and\ \bibinfo {author}
  {\bibfnamefont {F.~W.}\ \bibnamefont {Starr}},\ }\href
  {\doibase10.1039/B000058M} {\bibfield  {journal} {\bibinfo  {journal} {Phys.
  Chem. Chem. Phys.}\ }\textbf {\bibinfo {volume} {2}},\ \bibinfo {pages}
  {1551} (\bibinfo {year} {2000})}\BibitemShut {NoStop}%
\bibitem [{\citenamefont {Stokely}\ \emph {et~al.}(2010)\citenamefont
  {Stokely}, \citenamefont {Mazza}, \citenamefont {Stanley},\ and\
  \citenamefont {Franzese}}]{stokely2010}%
  \BibitemOpen
  \bibfield  {author} {\bibinfo {author} {\bibfnamefont {K.}~\bibnamefont
  {Stokely}}, \bibinfo {author} {\bibfnamefont {M.~G.}\ \bibnamefont {Mazza}},
  \bibinfo {author} {\bibfnamefont {H.~E.}\ \bibnamefont {Stanley}}, \ and\
  \bibinfo {author} {\bibfnamefont {G.}~\bibnamefont {Franzese}},\ }\href@noop
  {} {\bibfield  {journal} {\bibinfo  {journal} {Proc. Natl. Acad. Sci.
  U.S.A.}\ }\textbf {\bibinfo {volume} {107}},\ \bibinfo {pages} {1301}
  (\bibinfo {year} {2010})}\BibitemShut {NoStop}%
\bibitem [{\citenamefont {Tanaka}(2000)}]{tanaka2000EPL}%
  \BibitemOpen
  \bibfield  {author} {\bibinfo {author} {\bibfnamefont {H.}~\bibnamefont
  {Tanaka}},\ }\href@noop {} {\bibfield  {journal} {\bibinfo  {journal}
  {Europhys. Lett.}\ }\textbf {\bibinfo {volume} {50}},\ \bibinfo {pages} {340}
  (\bibinfo {year} {2000})}\BibitemShut {NoStop}%
\bibitem [{\citenamefont {Mishima}(2010{\natexlab{b}})}]{mishima2010review}%
  \BibitemOpen
  \bibfield  {author} {\bibinfo {author} {\bibfnamefont {O.}~\bibnamefont
  {Mishima}},\ }\href {\doibase10.2183/pjab.86.165} {\bibfield  {journal}
  {\bibinfo  {journal} {Proc. Jpn. Acad., Ser. B}\ }\textbf {\bibinfo {volume}
  {86}},\ \bibinfo {pages} {165} (\bibinfo {year}
  {2010}{\natexlab{b}})}\BibitemShut {NoStop}%
\bibitem [{\citenamefont {Mishima}\ and\ \citenamefont
  {Stanley}(1998{\natexlab{b}})}]{mishima1998}%
  \BibitemOpen
  \bibfield  {author} {\bibinfo {author} {\bibfnamefont {O.}~\bibnamefont
  {Mishima}}\ and\ \bibinfo {author} {\bibfnamefont {H.~E.}\ \bibnamefont
  {Stanley}},\ }\href@noop {} {\bibfield  {journal} {\bibinfo  {journal}
  {Nature}\ }\textbf {\bibinfo {volume} {392}},\ \bibinfo {pages} {164}
  (\bibinfo {year} {1998}{\natexlab{b}})}\BibitemShut {NoStop}%
\bibitem [{\citenamefont {Mishima}(2000)}]{mishima2000}%
  \BibitemOpen
  \bibfield  {author} {\bibinfo {author} {\bibfnamefont {O.}~\bibnamefont
  {Mishima}},\ }\href@noop {} {\bibfield  {journal} {\bibinfo  {journal} {Phys.
  Rev. Lett.}\ }\textbf {\bibinfo {volume} {85}},\ \bibinfo {pages} {334}
  (\bibinfo {year} {2000})}\BibitemShut {NoStop}%
\bibitem [{\citenamefont {Mishima}(2011)}]{mishima2011}%
  \BibitemOpen
  \bibfield  {author} {\bibinfo {author} {\bibfnamefont {O.}~\bibnamefont
  {Mishima}},\ }\href@noop {} {\bibfield  {journal} {\bibinfo  {journal}
  {J.~Phys. Chem.~B}\ }\textbf {\bibinfo {volume} {115}},\ \bibinfo {pages}
  {14064} (\bibinfo {year} {2011})}\BibitemShut {NoStop}%
\bibitem [{\citenamefont {Amann-Winkel}\ \emph {et~al.}(2013)\citenamefont
  {Amann-Winkel}, \citenamefont {Gainaru}, \citenamefont {Handle},
  \citenamefont {Seidl}, \citenamefont {Nelson}, \citenamefont {Böhmer},\ and\
  \citenamefont {Loerting}}]{amannwinkel2013}%
  \BibitemOpen
  \bibfield  {author} {\bibinfo {author} {\bibfnamefont {K.}~\bibnamefont
  {Amann-Winkel}}, \bibinfo {author} {\bibfnamefont {C.}~\bibnamefont
  {Gainaru}}, \bibinfo {author} {\bibfnamefont {P.~H.}\ \bibnamefont {Handle}},
  \bibinfo {author} {\bibfnamefont {M.}~\bibnamefont {Seidl}}, \bibinfo
  {author} {\bibfnamefont {H.}~\bibnamefont {Nelson}}, \bibinfo {author}
  {\bibfnamefont {R.}~\bibnamefont {Böhmer}}, \ and\ \bibinfo {author}
  {\bibfnamefont {T.}~\bibnamefont {Loerting}},\ }\href
  {\doibase10.1073/pnas.1311718110} {\bibfield  {journal} {\bibinfo  {journal}
  {Proc. Natl. Acad. Sci. U.S.A.}\ }\textbf {\bibinfo {volume} {110}},\
  \bibinfo {pages} {17720} (\bibinfo {year} {2013})}\BibitemShut {NoStop}%
\bibitem [{\citenamefont {{General Discussion}}(2013)}]{discussion2013A}%
  \BibitemOpen
  \bibfield  {author} {\bibinfo {author} {\bibnamefont {{General
  Discussion}}},\ }\href {\doibase10.1039/C3FD90036C} {\bibfield  {journal}
  {\bibinfo  {journal} {Faraday Discuss.}\ }\textbf {\bibinfo {volume} {167}},\
  \bibinfo {pages} {109} (\bibinfo {year} {2013})}\BibitemShut {NoStop}%
\bibitem [{\citenamefont {Stillinger}\ and\ \citenamefont
  {Rahman}(1974)}]{stillinger1974}%
  \BibitemOpen
  \bibfield  {author} {\bibinfo {author} {\bibfnamefont {F.~H.}\ \bibnamefont
  {Stillinger}}\ and\ \bibinfo {author} {\bibfnamefont {A.}~\bibnamefont
  {Rahman}},\ }\href@noop {} {\bibfield  {journal} {\bibinfo  {journal}
  {J.~Chem. Phys.}\ }\textbf {\bibinfo {volume} {60}},\ \bibinfo {pages} {1545}
  (\bibinfo {year} {1974})}\BibitemShut {NoStop}%
\bibitem [{\citenamefont {Liu}, \citenamefont {Panagiotopoulos},\ and\
  \citenamefont {Debenedetti}(2009)}]{liu2009}%
  \BibitemOpen
  \bibfield  {author} {\bibinfo {author} {\bibfnamefont {Y.}~\bibnamefont
  {Liu}}, \bibinfo {author} {\bibfnamefont {A.~Z.}\ \bibnamefont
  {Panagiotopoulos}}, \ and\ \bibinfo {author} {\bibfnamefont {P.~G.}\
  \bibnamefont {Debenedetti}},\ }\href@noop {} {\bibfield  {journal} {\bibinfo
  {journal} {J.~Chem. Phys.}\ }\textbf {\bibinfo {volume} {131}},\ \bibinfo
  {pages} {104508} (\bibinfo {year} {2009})}\BibitemShut {NoStop}%
\bibitem [{\citenamefont {Sciortino}, \citenamefont {Saika-Voivod},\ and\
  \citenamefont {Poole}(2011)}]{sciortino2011}%
  \BibitemOpen
  \bibfield  {author} {\bibinfo {author} {\bibfnamefont {F.}~\bibnamefont
  {Sciortino}}, \bibinfo {author} {\bibfnamefont {I.}~\bibnamefont
  {Saika-Voivod}}, \ and\ \bibinfo {author} {\bibfnamefont {P.~H.}\
  \bibnamefont {Poole}},\ }\href@noop {} {\bibfield  {journal} {\bibinfo
  {journal} {Phys. Chem. Chem. Phys.}\ }\textbf {\bibinfo {volume} {13}},\
  \bibinfo {pages} {19759} (\bibinfo {year} {2011})}\BibitemShut {NoStop}%
\bibitem [{\citenamefont {Liu}\ \emph {et~al.}(2012)\citenamefont {Liu},
  \citenamefont {Palmer}, \citenamefont {Panagiotopoulos},\ and\ \citenamefont
  {Debenedetti}}]{liu2012}%
  \BibitemOpen
  \bibfield  {author} {\bibinfo {author} {\bibfnamefont {Y.}~\bibnamefont
  {Liu}}, \bibinfo {author} {\bibfnamefont {J.~C.}\ \bibnamefont {Palmer}},
  \bibinfo {author} {\bibfnamefont {A.~Z.}\ \bibnamefont {Panagiotopoulos}}, \
  and\ \bibinfo {author} {\bibfnamefont {P.~G.}\ \bibnamefont {Debenedetti}},\
  }\href@noop {} {\bibfield  {journal} {\bibinfo  {journal} {J.~Chem. Phys.}\
  }\textbf {\bibinfo {volume} {137}},\ \bibinfo {pages} {214505} (\bibinfo
  {year} {2012})}\BibitemShut {NoStop}%
\bibitem [{\citenamefont {Kesselring}\ \emph {et~al.}(2012)\citenamefont
  {Kesselring}, \citenamefont {Franzese}, \citenamefont {Buldyrev},
  \citenamefont {Herrmann},\ and\ \citenamefont {Stanley}}]{kesselring2012}%
  \BibitemOpen
  \bibfield  {author} {\bibinfo {author} {\bibfnamefont {T.~A.}\ \bibnamefont
  {Kesselring}}, \bibinfo {author} {\bibfnamefont {G.}~\bibnamefont
  {Franzese}}, \bibinfo {author} {\bibfnamefont {S.~V.}\ \bibnamefont
  {Buldyrev}}, \bibinfo {author} {\bibfnamefont {H.~J.}\ \bibnamefont
  {Herrmann}}, \ and\ \bibinfo {author} {\bibfnamefont {H.~E.}\ \bibnamefont
  {Stanley}},\ }\href@noop {} {\bibfield  {journal} {\bibinfo  {journal} {Sci.
  Rep.}\ }\textbf {\bibinfo {volume} {2}},\ \bibinfo {pages} {474} (\bibinfo
  {year} {2012})}\BibitemShut {NoStop}%
\bibitem [{\citenamefont {Kesselring}\ \emph {et~al.}(2013)\citenamefont
  {Kesselring}, \citenamefont {Lascaris}, \citenamefont {Franzese},
  \citenamefont {Buldyrev}, \citenamefont {Herrmann},\ and\ \citenamefont
  {Stanley}}]{kesselring2013}%
  \BibitemOpen
  \bibfield  {author} {\bibinfo {author} {\bibfnamefont {T.~A.}\ \bibnamefont
  {Kesselring}}, \bibinfo {author} {\bibfnamefont {E.}~\bibnamefont
  {Lascaris}}, \bibinfo {author} {\bibfnamefont {G.}~\bibnamefont {Franzese}},
  \bibinfo {author} {\bibfnamefont {S.~V.}\ \bibnamefont {Buldyrev}}, \bibinfo
  {author} {\bibfnamefont {H.~J.}\ \bibnamefont {Herrmann}}, \ and\ \bibinfo
  {author} {\bibfnamefont {H.~E.}\ \bibnamefont {Stanley}},\ }\href@noop {}
  {\bibfield  {journal} {\bibinfo  {journal} {J.~Chem. Phys.}\ }\textbf
  {\bibinfo {volume} {138}},\ \bibinfo {pages} {244506} (\bibinfo {year}
  {2013})}\BibitemShut {NoStop}%
\bibitem [{\citenamefont {Cuthbertson}\ and\ \citenamefont
  {Poole}(2011)}]{cuthbertsonpoole2011}%
  \BibitemOpen
  \bibfield  {author} {\bibinfo {author} {\bibfnamefont {M.~J.}\ \bibnamefont
  {Cuthbertson}}\ and\ \bibinfo {author} {\bibfnamefont {P.~H.}\ \bibnamefont
  {Poole}},\ }\href@noop {} {\bibfield  {journal} {\bibinfo  {journal} {Phys.
  Rev. Lett.}\ }\textbf {\bibinfo {volume} {106}},\ \bibinfo {pages} {115706}
  (\bibinfo {year} {2011})}\BibitemShut {NoStop}%
\bibitem [{\citenamefont {Poole}\ \emph {et~al.}(2013)\citenamefont {Poole},
  \citenamefont {Bowles}, \citenamefont {Saika-Voivod},\ and\ \citenamefont
  {Sciortino}}]{poole2013}%
  \BibitemOpen
  \bibfield  {author} {\bibinfo {author} {\bibfnamefont {P.~H.}\ \bibnamefont
  {Poole}}, \bibinfo {author} {\bibfnamefont {R.~K.}\ \bibnamefont {Bowles}},
  \bibinfo {author} {\bibfnamefont {I.}~\bibnamefont {Saika-Voivod}}, \ and\
  \bibinfo {author} {\bibfnamefont {F.}~\bibnamefont {Sciortino}},\ }\href@noop
  {} {\bibfield  {journal} {\bibinfo  {journal} {J.~Chem. Phys.}\ }\textbf
  {\bibinfo {volume} {138}},\ \bibinfo {pages} {034505} (\bibinfo {year}
  {2013})}\BibitemShut {NoStop}%
\bibitem [{\citenamefont {Palmer}, \citenamefont {Car},\ and\ \citenamefont
  {Debenedetti}(2013)}]{palmer2013}%
  \BibitemOpen
  \bibfield  {author} {\bibinfo {author} {\bibfnamefont {J.~C.}\ \bibnamefont
  {Palmer}}, \bibinfo {author} {\bibfnamefont {R.}~\bibnamefont {Car}}, \ and\
  \bibinfo {author} {\bibfnamefont {P.~G.}\ \bibnamefont {Debenedetti}},\
  }\href {\doibase10.1039/C3FD00074E} {\bibfield  {journal} {\bibinfo
  {journal} {Faraday Discuss.}\ }\textbf {\bibinfo {volume} {167}},\ \bibinfo
  {pages} {77} (\bibinfo {year} {2013})}\BibitemShut {NoStop}%
\bibitem [{\citenamefont {Steinhardt}, \citenamefont {Nelson},\ and\
  \citenamefont {Ronchetti}(1983)}]{steinhardt1983}%
  \BibitemOpen
  \bibfield  {author} {\bibinfo {author} {\bibfnamefont {P.~J.}\ \bibnamefont
  {Steinhardt}}, \bibinfo {author} {\bibfnamefont {D.~R.}\ \bibnamefont
  {Nelson}}, \ and\ \bibinfo {author} {\bibfnamefont {M.}~\bibnamefont
  {Ronchetti}},\ }\href {\doibase10.1103/PhysRevB.28.784} {\bibfield  {journal}
  {\bibinfo  {journal} {Phys. Rev. B}\ }\textbf {\bibinfo {volume} {28}},\
  \bibinfo {pages} {784} (\bibinfo {year} {1983})}\BibitemShut {NoStop}%
\bibitem [{\citenamefont {Limmer}\ and\ \citenamefont
  {Chandler}(2011)}]{limmer2011}%
  \BibitemOpen
  \bibfield  {author} {\bibinfo {author} {\bibfnamefont {D.~T.}\ \bibnamefont
  {Limmer}}\ and\ \bibinfo {author} {\bibfnamefont {D.}~\bibnamefont
  {Chandler}},\ }\href@noop {} {\bibfield  {journal} {\bibinfo  {journal}
  {J.~Chem. Phys.}\ }\textbf {\bibinfo {volume} {135}},\ \bibinfo {pages}
  {134503} (\bibinfo {year} {2011})}\BibitemShut {NoStop}%
\bibitem [{\citenamefont {Limmer}\ and\ \citenamefont
  {Chandler}(2013)}]{limmer2013}%
  \BibitemOpen
  \bibfield  {author} {\bibinfo {author} {\bibfnamefont {D.~T.}\ \bibnamefont
  {Limmer}}\ and\ \bibinfo {author} {\bibfnamefont {D.}~\bibnamefont
  {Chandler}},\ }\href@noop {} {\bibfield  {journal} {\bibinfo  {journal}
  {J.~Chem. Phys.}\ }\textbf {\bibinfo {volume} {138}},\ \bibinfo {pages}
  {214504} (\bibinfo {year} {2013})}\BibitemShut {NoStop}%
\bibitem [{\citenamefont {English}, \citenamefont {Kusalik},\ and\
  \citenamefont {Tse}(2013)}]{english2013}%
  \BibitemOpen
  \bibfield  {author} {\bibinfo {author} {\bibfnamefont {N.~J.}\ \bibnamefont
  {English}}, \bibinfo {author} {\bibfnamefont {P.~G.}\ \bibnamefont
  {Kusalik}}, \ and\ \bibinfo {author} {\bibfnamefont {J.~S.}\ \bibnamefont
  {Tse}},\ }\href {\doibase10.1063/1.4818876} {\bibfield  {journal} {\bibinfo
  {journal} {J.~Chem. Phys.}\ }\textbf {\bibinfo {volume} {139}},\ \bibinfo
  {pages} {084508} (\bibinfo {year} {2013})}\BibitemShut {NoStop}%
\bibitem [{\citenamefont {Molinero}\ and\ \citenamefont
  {Moore}(2009)}]{molinero2009}%
  \BibitemOpen
  \bibfield  {author} {\bibinfo {author} {\bibfnamefont {V.}~\bibnamefont
  {Molinero}}\ and\ \bibinfo {author} {\bibfnamefont {E.~B.}\ \bibnamefont
  {Moore}},\ }\href@noop {} {\bibfield  {journal} {\bibinfo  {journal}
  {J.~Phys. Chem.~B}\ }\textbf {\bibinfo {volume} {113}},\ \bibinfo {pages}
  {4008} (\bibinfo {year} {2009})}\BibitemShut {NoStop}%
\bibitem [{\citenamefont {Moore}\ and\ \citenamefont
  {Molinero}(2011)}]{moore2011}%
  \BibitemOpen
  \bibfield  {author} {\bibinfo {author} {\bibfnamefont {E.~B.}\ \bibnamefont
  {Moore}}\ and\ \bibinfo {author} {\bibfnamefont {V.}~\bibnamefont
  {Molinero}},\ }\href@noop {} {\bibfield  {journal} {\bibinfo  {journal}
  {Nature}\ }\textbf {\bibinfo {volume} {479}},\ \bibinfo {pages} {506}
  (\bibinfo {year} {2011})}\BibitemShut {NoStop}%
\bibitem [{\citenamefont {Moore}\ and\ \citenamefont
  {Molinero}(2009)}]{moore2009}%
  \BibitemOpen
  \bibfield  {author} {\bibinfo {author} {\bibfnamefont {E.~B.}\ \bibnamefont
  {Moore}}\ and\ \bibinfo {author} {\bibfnamefont {V.}~\bibnamefont
  {Molinero}},\ }\href@noop {} {\bibfield  {journal} {\bibinfo  {journal}
  {J.~Chem. Phys.}\ }\textbf {\bibinfo {volume} {130}},\ \bibinfo {pages}
  {244505} (\bibinfo {year} {2009})}\BibitemShut {NoStop}%
\bibitem [{\citenamefont {Nilsson}, \citenamefont {Huang},\ and\ \citenamefont
  {Pettersson}(2012)}]{nilsson2012}%
  \BibitemOpen
  \bibfield  {author} {\bibinfo {author} {\bibfnamefont {A.}~\bibnamefont
  {Nilsson}}, \bibinfo {author} {\bibfnamefont {C.}~\bibnamefont {Huang}}, \
  and\ \bibinfo {author} {\bibfnamefont {L.~G.}\ \bibnamefont {Pettersson}},\
  }\href@noop {} {\bibfield  {journal} {\bibinfo  {journal} {J. Mol. Liq.}\
  }\textbf {\bibinfo {volume} {176}},\ \bibinfo {pages} {2} (\bibinfo {year}
  {2012})}\BibitemShut {NoStop}%
\bibitem [{\citenamefont {Taschin}\ \emph {et~al.}(2013)\citenamefont
  {Taschin}, \citenamefont {Bartolini}, \citenamefont {Eramo}, \citenamefont
  {Righini},\ and\ \citenamefont {Torre}}]{taschin2013}%
  \BibitemOpen
  \bibfield  {author} {\bibinfo {author} {\bibfnamefont {A.}~\bibnamefont
  {Taschin}}, \bibinfo {author} {\bibfnamefont {P.}~\bibnamefont {Bartolini}},
  \bibinfo {author} {\bibfnamefont {R.}~\bibnamefont {Eramo}}, \bibinfo
  {author} {\bibfnamefont {R.}~\bibnamefont {Righini}}, \ and\ \bibinfo
  {author} {\bibfnamefont {R.}~\bibnamefont {Torre}},\ }\href@noop {}
  {\bibfield  {journal} {\bibinfo  {journal} {Nat. Commun.}\ }\textbf {\bibinfo
  {volume} {4}},\ \bibinfo {pages} {2401} (\bibinfo {year} {2013})}\BibitemShut
  {NoStop}%
\bibitem [{\citenamefont {Wikfeldt}, \citenamefont {Nilsson},\ and\
  \citenamefont {Pettersson}(2011)}]{wikfeldt2011}%
  \BibitemOpen
  \bibfield  {author} {\bibinfo {author} {\bibfnamefont {K.~T.}\ \bibnamefont
  {Wikfeldt}}, \bibinfo {author} {\bibfnamefont {A.}~\bibnamefont {Nilsson}}, \
  and\ \bibinfo {author} {\bibfnamefont {L.~G.~M.}\ \bibnamefont
  {Pettersson}},\ }\href@noop {} {\bibfield  {journal} {\bibinfo  {journal}
  {Phys. Chem. Chem. Phys.}\ }\textbf {\bibinfo {volume} {13}},\ \bibinfo
  {pages} {19918} (\bibinfo {year} {2011})}\BibitemShut {NoStop}%
\bibitem [{\citenamefont {Holten}\ and\ \citenamefont
  {Anisimov}(2012{\natexlab{a}})}]{holtentwostate}%
  \BibitemOpen
  \bibfield  {author} {\bibinfo {author} {\bibfnamefont {V.}~\bibnamefont
  {Holten}}\ and\ \bibinfo {author} {\bibfnamefont {M.~A.}\ \bibnamefont
  {Anisimov}},\ }\href@noop {} {\bibfield  {journal} {\bibinfo  {journal} {Sci.
  Rep.}\ }\textbf {\bibinfo {volume} {2}},\ \bibinfo {pages} {713} (\bibinfo
  {year} {2012}{\natexlab{a}})}\BibitemShut {NoStop}%
\bibitem [{\citenamefont {Tokushima}\ \emph {et~al.}(2008)\citenamefont
  {Tokushima}, \citenamefont {Harada}, \citenamefont {Takahashi}, \citenamefont
  {Senba}, \citenamefont {Ohashi}, \citenamefont {Pettersson}, \citenamefont
  {Nilsson},\ and\ \citenamefont {Shin}}]{tokushima2008}%
  \BibitemOpen
  \bibfield  {author} {\bibinfo {author} {\bibfnamefont {T.}~\bibnamefont
  {Tokushima}}, \bibinfo {author} {\bibfnamefont {Y.}~\bibnamefont {Harada}},
  \bibinfo {author} {\bibfnamefont {O.}~\bibnamefont {Takahashi}}, \bibinfo
  {author} {\bibfnamefont {Y.}~\bibnamefont {Senba}}, \bibinfo {author}
  {\bibfnamefont {H.}~\bibnamefont {Ohashi}}, \bibinfo {author} {\bibfnamefont
  {L.~G.~M.}\ \bibnamefont {Pettersson}}, \bibinfo {author} {\bibfnamefont
  {A.}~\bibnamefont {Nilsson}}, \ and\ \bibinfo {author} {\bibfnamefont
  {S.}~\bibnamefont {Shin}},\ }\href {\doibase10.1016/j.cplett.2008.04.077}
  {\bibfield  {journal} {\bibinfo  {journal} {Chem. Phys. Lett.}\ }\textbf
  {\bibinfo {volume} {460}},\ \bibinfo {pages} {387} (\bibinfo {year}
  {2008})}\BibitemShut {NoStop}%
\bibitem [{\citenamefont {Wernet}\ \emph {et~al.}(2004)\citenamefont {Wernet},
  \citenamefont {Nordlund}, \citenamefont {Bergmann}, \citenamefont
  {Cavalleri}, \citenamefont {Odelius}, \citenamefont {Ogasawara},
  \citenamefont {Näslund}, \citenamefont {Hirsch}, \citenamefont {Ojamäe},
  \citenamefont {Glatzel}, \citenamefont {Pettersson},\ and\ \citenamefont
  {Nilsson}}]{wernet2004}%
  \BibitemOpen
  \bibfield  {author} {\bibinfo {author} {\bibfnamefont {{\relax
  Ph}.}~\bibnamefont {Wernet}}, \bibinfo {author} {\bibfnamefont
  {D.}~\bibnamefont {Nordlund}}, \bibinfo {author} {\bibfnamefont
  {U.}~\bibnamefont {Bergmann}}, \bibinfo {author} {\bibfnamefont
  {M.}~\bibnamefont {Cavalleri}}, \bibinfo {author} {\bibfnamefont
  {M.}~\bibnamefont {Odelius}}, \bibinfo {author} {\bibfnamefont
  {H.}~\bibnamefont {Ogasawara}}, \bibinfo {author} {\bibfnamefont {L.~{\AA}.}\
  \bibnamefont {Näslund}}, \bibinfo {author} {\bibfnamefont {T.~K.}\
  \bibnamefont {Hirsch}}, \bibinfo {author} {\bibfnamefont {L.}~\bibnamefont
  {Ojamäe}}, \bibinfo {author} {\bibfnamefont {P.}~\bibnamefont {Glatzel}},
  \bibinfo {author} {\bibfnamefont {L.~G.~M.}\ \bibnamefont {Pettersson}}, \
  and\ \bibinfo {author} {\bibfnamefont {A.}~\bibnamefont {Nilsson}},\ }\href
  {\doibase10.1126/science.1096205} {\bibfield  {journal} {\bibinfo  {journal}
  {Science}\ }\textbf {\bibinfo {volume} {304}},\ \bibinfo {pages} {995}
  (\bibinfo {year} {2004})}\BibitemShut {NoStop}%
\bibitem [{\citenamefont {Huang}\ \emph {et~al.}(2009)\citenamefont {Huang},
  \citenamefont {Wikfeldt}, \citenamefont {Tokushima}, \citenamefont
  {Nordlund}, \citenamefont {Harada}, \citenamefont {Bergmann}, \citenamefont
  {Niebuhr}, \citenamefont {Weiss}, \citenamefont {Horikawa}, \citenamefont
  {Leetmaa}, \citenamefont {Ljungberg}, \citenamefont {Takahashi},
  \citenamefont {Lenz}, \citenamefont {Ojamäe}, \citenamefont {Lyubartsev},
  \citenamefont {Shin}, \citenamefont {Pettersson},\ and\ \citenamefont
  {Nilsson}}]{huang2009}%
  \BibitemOpen
  \bibfield  {author} {\bibinfo {author} {\bibfnamefont {C.}~\bibnamefont
  {Huang}}, \bibinfo {author} {\bibfnamefont {K.~T.}\ \bibnamefont {Wikfeldt}},
  \bibinfo {author} {\bibfnamefont {T.}~\bibnamefont {Tokushima}}, \bibinfo
  {author} {\bibfnamefont {D.}~\bibnamefont {Nordlund}}, \bibinfo {author}
  {\bibfnamefont {Y.}~\bibnamefont {Harada}}, \bibinfo {author} {\bibfnamefont
  {U.}~\bibnamefont {Bergmann}}, \bibinfo {author} {\bibfnamefont
  {M.}~\bibnamefont {Niebuhr}}, \bibinfo {author} {\bibfnamefont {T.~M.}\
  \bibnamefont {Weiss}}, \bibinfo {author} {\bibfnamefont {Y.}~\bibnamefont
  {Horikawa}}, \bibinfo {author} {\bibfnamefont {M.}~\bibnamefont {Leetmaa}},
  \bibinfo {author} {\bibfnamefont {M.~P.}\ \bibnamefont {Ljungberg}}, \bibinfo
  {author} {\bibfnamefont {O.}~\bibnamefont {Takahashi}}, \bibinfo {author}
  {\bibfnamefont {A.}~\bibnamefont {Lenz}}, \bibinfo {author} {\bibfnamefont
  {L.}~\bibnamefont {Ojamäe}}, \bibinfo {author} {\bibfnamefont {A.~P.}\
  \bibnamefont {Lyubartsev}}, \bibinfo {author} {\bibfnamefont
  {S.}~\bibnamefont {Shin}}, \bibinfo {author} {\bibfnamefont {L.~G.~M.}\
  \bibnamefont {Pettersson}}, \ and\ \bibinfo {author} {\bibfnamefont
  {A.}~\bibnamefont {Nilsson}},\ }\href {\doibase10.1073/pnas.0904743106}
  {\bibfield  {journal} {\bibinfo  {journal} {Proc. Natl. Acad. Sci. U.S.A.}\
  }\textbf {\bibinfo {volume} {106}},\ \bibinfo {pages} {15214} (\bibinfo
  {year} {2009})}\BibitemShut {NoStop}%
\bibitem [{\citenamefont {Xu}\ \emph {et~al.}(2009)\citenamefont {Xu},
  \citenamefont {Mallamace}, \citenamefont {Yan}, \citenamefont {Starr},
  \citenamefont {Buldyrev},\ and\ \citenamefont {Stanley}}]{xu2009}%
  \BibitemOpen
  \bibfield  {author} {\bibinfo {author} {\bibfnamefont {L.}~\bibnamefont
  {Xu}}, \bibinfo {author} {\bibfnamefont {F.}~\bibnamefont {Mallamace}},
  \bibinfo {author} {\bibfnamefont {Z.}~\bibnamefont {Yan}}, \bibinfo {author}
  {\bibfnamefont {F.~W.}\ \bibnamefont {Starr}}, \bibinfo {author}
  {\bibfnamefont {S.~V.}\ \bibnamefont {Buldyrev}}, \ and\ \bibinfo {author}
  {\bibfnamefont {H.~E.}\ \bibnamefont {Stanley}},\ }\href
  {\doibase10.1038/nphys1328} {\bibfield  {journal} {\bibinfo  {journal}
  {Nature Phys.}\ }\textbf {\bibinfo {volume} {5}},\ \bibinfo {pages} {565}
  (\bibinfo {year} {2009})}\BibitemShut {NoStop}%
\bibitem [{\citenamefont {Holten}\ \emph {et~al.}(2013)\citenamefont {Holten},
  \citenamefont {Limmer}, \citenamefont {Molinero},\ and\ \citenamefont
  {Anisimov}}]{holtenmW}%
  \BibitemOpen
  \bibfield  {author} {\bibinfo {author} {\bibfnamefont {V.}~\bibnamefont
  {Holten}}, \bibinfo {author} {\bibfnamefont {D.~T.}\ \bibnamefont {Limmer}},
  \bibinfo {author} {\bibfnamefont {V.}~\bibnamefont {Molinero}}, \ and\
  \bibinfo {author} {\bibfnamefont {M.~A.}\ \bibnamefont {Anisimov}},\
  }\href@noop {} {\bibfield  {journal} {\bibinfo  {journal} {J.~Chem. Phys.}\
  }\textbf {\bibinfo {volume} {138}},\ \bibinfo {pages} {174501} (\bibinfo
  {year} {2013})}\BibitemShut {NoStop}%
\bibitem [{\citenamefont {Bertrand}\ and\ \citenamefont
  {Anisimov}(2011)}]{bertrand2011}%
  \BibitemOpen
  \bibfield  {author} {\bibinfo {author} {\bibfnamefont {C.~E.}\ \bibnamefont
  {Bertrand}}\ and\ \bibinfo {author} {\bibfnamefont {M.~A.}\ \bibnamefont
  {Anisimov}},\ }\href@noop {} {\bibfield  {journal} {\bibinfo  {journal}
  {J.~Phys. Chem.~B}\ }\textbf {\bibinfo {volume} {115}},\ \bibinfo {pages}
  {14099} (\bibinfo {year} {2011})}\BibitemShut {NoStop}%
\bibitem [{\citenamefont {Tanaka}(2013)}]{tanaka2013}%
  \BibitemOpen
  \bibfield  {author} {\bibinfo {author} {\bibfnamefont {H.}~\bibnamefont
  {Tanaka}},\ }\href {\doibase10.1039/c3fd00110e} {\bibfield  {journal}
  {\bibinfo  {journal} {Faraday Discuss.}\ }\textbf {\bibinfo {volume} {167}},\
  \bibinfo {pages} {9} (\bibinfo {year} {2013})}\BibitemShut {NoStop}%
\bibitem [{\citenamefont {Tu}\ \emph {et~al.}(2012)\citenamefont {Tu},
  \citenamefont {Buldyrev}, \citenamefont {Liu}, \citenamefont {Fang},\ and\
  \citenamefont {Stanley}}]{tu2012}%
  \BibitemOpen
  \bibfield  {author} {\bibinfo {author} {\bibfnamefont {Y.}~\bibnamefont
  {Tu}}, \bibinfo {author} {\bibfnamefont {S.~V.}\ \bibnamefont {Buldyrev}},
  \bibinfo {author} {\bibfnamefont {Z.}~\bibnamefont {Liu}}, \bibinfo {author}
  {\bibfnamefont {H.}~\bibnamefont {Fang}}, \ and\ \bibinfo {author}
  {\bibfnamefont {H.~E.}\ \bibnamefont {Stanley}},\ }\href@noop {} {\bibfield
  {journal} {\bibinfo  {journal} {EPL}\ }\textbf {\bibinfo {volume} {97}},\
  \bibinfo {pages} {56005} (\bibinfo {year} {2012})}\BibitemShut {NoStop}%
\bibitem [{\citenamefont {Xu}\ \emph {et~al.}(2006)\citenamefont {Xu},
  \citenamefont {Buldyrev}, \citenamefont {Angell},\ and\ \citenamefont
  {Stanley}}]{xu2006}%
  \BibitemOpen
  \bibfield  {author} {\bibinfo {author} {\bibfnamefont {L.}~\bibnamefont
  {Xu}}, \bibinfo {author} {\bibfnamefont {S.~V.}\ \bibnamefont {Buldyrev}},
  \bibinfo {author} {\bibfnamefont {C.~A.}\ \bibnamefont {Angell}}, \ and\
  \bibinfo {author} {\bibfnamefont {H.~E.}\ \bibnamefont {Stanley}},\
  }\href@noop {} {\bibfield  {journal} {\bibinfo  {journal} {Phys. Rev. E}\
  }\textbf {\bibinfo {volume} {74}},\ \bibinfo {pages} {031108} (\bibinfo
  {year} {2006})}\BibitemShut {NoStop}%
\bibitem [{\citenamefont {Whiting}(1884)}]{whiting1884}%
  \BibitemOpen
  \bibfield  {author} {\bibinfo {author} {\bibfnamefont {H.}~\bibnamefont
  {Whiting}},\ }\href@noop {} {\bibfield  {journal} {\bibinfo  {journal} {Proc.
  Am. Acad. Arts Sci.}\ }\textbf {\bibinfo {volume} {19}},\ \bibinfo {pages}
  {353} (\bibinfo {year} {1884})}\BibitemShut {NoStop}%
\bibitem [{\citenamefont {R{\"o}ntgen}(1892)}]{roentgen}%
  \BibitemOpen
  \bibfield  {author} {\bibinfo {author} {\bibfnamefont {W.~C.}\ \bibnamefont
  {R{\"o}ntgen}},\ }\href@noop {} {\bibfield  {journal} {\bibinfo  {journal}
  {Ann. Phys. (Leipzig)}\ }\textbf {\bibinfo {volume} {281}},\ \bibinfo {pages}
  {91} (\bibinfo {year} {1892})}\BibitemShut {NoStop}%
\bibitem [{\citenamefont {McMillan}(2004)}]{mcmillan2004}%
  \BibitemOpen
  \bibfield  {author} {\bibinfo {author} {\bibfnamefont {P.~F.}\ \bibnamefont
  {McMillan}},\ }\href@noop {} {\bibfield  {journal} {\bibinfo  {journal} {J.
  Mater. Chem.}\ }\textbf {\bibinfo {volume} {14}},\ \bibinfo {pages} {1506}
  (\bibinfo {year} {2004})}\BibitemShut {NoStop}%
\bibitem [{\citenamefont {Wilding}, \citenamefont {Wilson},\ and\ \citenamefont
  {McMillan}(2006)}]{mcmillan2006}%
  \BibitemOpen
  \bibfield  {author} {\bibinfo {author} {\bibfnamefont {M.~C.}\ \bibnamefont
  {Wilding}}, \bibinfo {author} {\bibfnamefont {M.}~\bibnamefont {Wilson}}, \
  and\ \bibinfo {author} {\bibfnamefont {P.~F.}\ \bibnamefont {McMillan}},\
  }\href@noop {} {\bibfield  {journal} {\bibinfo  {journal} {Chem. Soc. Rev.}\
  }\textbf {\bibinfo {volume} {35}},\ \bibinfo {pages} {964} (\bibinfo {year}
  {2006})}\BibitemShut {NoStop}%
\bibitem [{\citenamefont {Vedamuthu}, \citenamefont {Singh},\ and\
  \citenamefont {Robinson}(1994)}]{vedamuthu1994}%
  \BibitemOpen
  \bibfield  {author} {\bibinfo {author} {\bibfnamefont {M.}~\bibnamefont
  {Vedamuthu}}, \bibinfo {author} {\bibfnamefont {S.}~\bibnamefont {Singh}}, \
  and\ \bibinfo {author} {\bibfnamefont {G.~W.}\ \bibnamefont {Robinson}},\
  }\href@noop {} {\bibfield  {journal} {\bibinfo  {journal} {J. Phys. Chem.}\
  }\textbf {\bibinfo {volume} {98}},\ \bibinfo {pages} {2222} (\bibinfo {year}
  {1994})}\BibitemShut {NoStop}%
\bibitem [{\citenamefont {Ponyatovsky}, \citenamefont {Sinitsyn},\ and\
  \citenamefont {Pozdnyakova}(1998)}]{ponyatovsky1998}%
  \BibitemOpen
  \bibfield  {author} {\bibinfo {author} {\bibfnamefont {E.~G.}\ \bibnamefont
  {Ponyatovsky}}, \bibinfo {author} {\bibfnamefont {V.~V.}\ \bibnamefont
  {Sinitsyn}}, \ and\ \bibinfo {author} {\bibfnamefont {T.~A.}\ \bibnamefont
  {Pozdnyakova}},\ }\href@noop {} {\bibfield  {journal} {\bibinfo  {journal}
  {J.~Chem. Phys.}\ }\textbf {\bibinfo {volume} {109}},\ \bibinfo {pages}
  {2413} (\bibinfo {year} {1998})}\BibitemShut {NoStop}%
\bibitem [{\citenamefont {Moynihan}(1996)}]{moynihan1997}%
  \BibitemOpen
  \bibfield  {author} {\bibinfo {author} {\bibfnamefont {C.~T.}\ \bibnamefont
  {Moynihan}},\ }\href@noop {} {\bibfield  {journal} {\bibinfo  {journal} {Mat.
  Res. Soc. Symp. Proc.}\ }\textbf {\bibinfo {volume} {455}},\ \bibinfo {pages}
  {411} (\bibinfo {year} {1996})}\BibitemShut {NoStop}%
\bibitem [{\citenamefont {Prigogine}\ and\ \citenamefont
  {Defay}(1954)}]{prigogine1954}%
  \BibitemOpen
  \bibfield  {author} {\bibinfo {author} {\bibfnamefont {I.}~\bibnamefont
  {Prigogine}}\ and\ \bibinfo {author} {\bibfnamefont {R.}~\bibnamefont
  {Defay}},\ }\href@noop {} {\emph {\bibinfo {title} {Chemical
  Thermodynamics}}}\ (\bibinfo  {publisher} {Longmans, Green \& Co.},\ \bibinfo
  {address} {London},\ \bibinfo {year} {1954})\BibitemShut {NoStop}%
\bibitem [{\citenamefont {Xu}\ \emph {et~al.}(2005)\citenamefont {Xu},
  \citenamefont {Kumar}, \citenamefont {Buldyrev}, \citenamefont {Chen},
  \citenamefont {Poole}, \citenamefont {Sciortino},\ and\ \citenamefont
  {Stanley}}]{xu2005}%
  \BibitemOpen
  \bibfield  {author} {\bibinfo {author} {\bibfnamefont {L.}~\bibnamefont
  {Xu}}, \bibinfo {author} {\bibfnamefont {P.}~\bibnamefont {Kumar}}, \bibinfo
  {author} {\bibfnamefont {S.~V.}\ \bibnamefont {Buldyrev}}, \bibinfo {author}
  {\bibfnamefont {S.-H.}\ \bibnamefont {Chen}}, \bibinfo {author}
  {\bibfnamefont {P.~H.}\ \bibnamefont {Poole}}, \bibinfo {author}
  {\bibfnamefont {F.}~\bibnamefont {Sciortino}}, \ and\ \bibinfo {author}
  {\bibfnamefont {H.~E.}\ \bibnamefont {Stanley}},\ }\href@noop {} {\bibfield
  {journal} {\bibinfo  {journal} {Proc. Natl. Acad. Sci. U.S.A.}\ }\textbf
  {\bibinfo {volume} {102}},\ \bibinfo {pages} {16558} (\bibinfo {year}
  {2005})}\BibitemShut {NoStop}%
\bibitem [{\citenamefont {Fuentevilla}\ and\ \citenamefont
  {Anisimov}(2006)}]{fuentevilla2006}%
  \BibitemOpen
  \bibfield  {author} {\bibinfo {author} {\bibfnamefont {D.~A.}\ \bibnamefont
  {Fuentevilla}}\ and\ \bibinfo {author} {\bibfnamefont {M.~A.}\ \bibnamefont
  {Anisimov}},\ }\href@noop {} {\bibfield  {journal} {\bibinfo  {journal}
  {Phys. Rev. Lett.}\ }\textbf {\bibinfo {volume} {97}},\ \bibinfo {pages}
  {195702} (\bibinfo {year} {2006})},\ \bibinfo {note} {erratum \textit{ibid.}
  \textbf{98}, 149904 (2007)}\BibitemShut {NoStop}%
\bibitem [{\citenamefont {Holten}\ \emph {et~al.}(2012)\citenamefont {Holten},
  \citenamefont {Bertrand}, \citenamefont {Anisimov},\ and\ \citenamefont
  {Sengers}}]{holtenSCW}%
  \BibitemOpen
  \bibfield  {author} {\bibinfo {author} {\bibfnamefont {V.}~\bibnamefont
  {Holten}}, \bibinfo {author} {\bibfnamefont {C.~E.}\ \bibnamefont
  {Bertrand}}, \bibinfo {author} {\bibfnamefont {M.~A.}\ \bibnamefont
  {Anisimov}}, \ and\ \bibinfo {author} {\bibfnamefont {J.~V.}\ \bibnamefont
  {Sengers}},\ }\href@noop {} {\bibfield  {journal} {\bibinfo  {journal}
  {J.~Chem. Phys.}\ }\textbf {\bibinfo {volume} {136}},\ \bibinfo {pages}
  {094507} (\bibinfo {year} {2012})}\BibitemShut {NoStop}%
\bibitem [{\citenamefont {Fisher}(1983)}]{fisher1983}%
  \BibitemOpen
  \bibfield  {author} {\bibinfo {author} {\bibfnamefont {M.~E.}\ \bibnamefont
  {Fisher}},\ }in\ \href@noop {} {\emph {\bibinfo {booktitle} {Critical
  Phenomena, Lecture Notes in Physics}}},\ Vol.\ \bibinfo {volume} {186},\
  \bibinfo {editor} {edited by\ \bibinfo {editor} {\bibfnamefont {F.~J.~W.}\
  \bibnamefont {Hahne}}}\ (\bibinfo  {publisher} {Springer},\ \bibinfo
  {address} {Berlin},\ \bibinfo {year} {1983})\ pp.\ \bibinfo {pages}
  {1--139}\BibitemShut {NoStop}%
\bibitem [{\citenamefont {Anisimov}\ and\ \citenamefont
  {Sengers}(2000)}]{anisimov2000inbook}%
  \BibitemOpen
  \bibfield  {author} {\bibinfo {author} {\bibfnamefont {M.~A.}\ \bibnamefont
  {Anisimov}}\ and\ \bibinfo {author} {\bibfnamefont {J.~V.}\ \bibnamefont
  {Sengers}},\ }in\ \href@noop {} {\emph {\bibinfo {booktitle} {Equations of
  State for Fluids and Fluid Mixtures}}},\ \bibinfo {series} {Experimental
  Thermodynamics}, Vol.~\bibinfo {volume} {V},\ \bibinfo {editor} {edited by\
  \bibinfo {editor} {\bibfnamefont {J.~V.}\ \bibnamefont {Sengers}}, \bibinfo
  {editor} {\bibfnamefont {R.~F.}\ \bibnamefont {Kayser}}, \bibinfo {editor}
  {\bibfnamefont {C.~J.}\ \bibnamefont {Peters}}, \ and\ \bibinfo {editor}
  {\bibfnamefont {H.~J.}\ \bibnamefont {White}, \bibfnamefont {Jr.}}}\
  (\bibinfo  {publisher} {Elsevier},\ \bibinfo {address} {Amsterdam},\ \bibinfo
  {year} {2000})\ Chap.~\bibinfo {chapter} {11}, pp.\ \bibinfo {pages}
  {381--434}\BibitemShut {NoStop}%
\bibitem [{\citenamefont {Behnejad}, \citenamefont {Sengers},\ and\
  \citenamefont {Anisimov}(2010)}]{behnejad2010}%
  \BibitemOpen
  \bibfield  {author} {\bibinfo {author} {\bibfnamefont {H.}~\bibnamefont
  {Behnejad}}, \bibinfo {author} {\bibfnamefont {J.~V.}\ \bibnamefont
  {Sengers}}, \ and\ \bibinfo {author} {\bibfnamefont {M.~A.}\ \bibnamefont
  {Anisimov}},\ }in\ \href@noop {} {\emph {\bibinfo {booktitle} {Applied
  Thermodynamics of Fluids}}},\ \bibinfo {editor} {edited by\ \bibinfo {editor}
  {\bibfnamefont {A.~R.~H.}\ \bibnamefont {Goodwin}}, \bibinfo {editor}
  {\bibfnamefont {J.~V.}\ \bibnamefont {Sengers}}, \ and\ \bibinfo {editor}
  {\bibfnamefont {C.~J.}\ \bibnamefont {Peters}}}\ (\bibinfo  {publisher} {RSC
  Publishing},\ \bibinfo {address} {Cambridge, UK},\ \bibinfo {year} {2010})\
  Chap.~\bibinfo {chapter} {10}, pp.\ \bibinfo {pages} {321--367}\BibitemShut
  {NoStop}%
\bibitem [{\citenamefont {Anisimov}\ \emph {et~al.}(1992)\citenamefont
  {Anisimov}, \citenamefont {Kiselev}, \citenamefont {Sengers},\ and\
  \citenamefont {Tang}}]{anisimov1992}%
  \BibitemOpen
  \bibfield  {author} {\bibinfo {author} {\bibfnamefont {M.~A.}\ \bibnamefont
  {Anisimov}}, \bibinfo {author} {\bibfnamefont {S.~B.}\ \bibnamefont
  {Kiselev}}, \bibinfo {author} {\bibfnamefont {J.~V.}\ \bibnamefont
  {Sengers}}, \ and\ \bibinfo {author} {\bibfnamefont {S.}~\bibnamefont
  {Tang}},\ }\href@noop {} {\bibfield  {journal} {\bibinfo  {journal} {Physica
  A}\ }\textbf {\bibinfo {volume} {188}},\ \bibinfo {pages} {487} (\bibinfo
  {year} {1992})}\BibitemShut {NoStop}%
\bibitem [{\citenamefont {Kim}\ \emph {et~al.}(2003)\citenamefont {Kim},
  \citenamefont {Anisimov}, \citenamefont {Sengers},\ and\ \citenamefont
  {Luijten}}]{kim2003b}%
  \BibitemOpen
  \bibfield  {author} {\bibinfo {author} {\bibfnamefont {Y.~C.}\ \bibnamefont
  {Kim}}, \bibinfo {author} {\bibfnamefont {M.~A.}\ \bibnamefont {Anisimov}},
  \bibinfo {author} {\bibfnamefont {J.~V.}\ \bibnamefont {Sengers}}, \ and\
  \bibinfo {author} {\bibfnamefont {E.}~\bibnamefont {Luijten}},\ }\href@noop
  {} {\bibfield  {journal} {\bibinfo  {journal} {J.~Stat. Phys.}\ }\textbf
  {\bibinfo {volume} {110}},\ \bibinfo {pages} {591} (\bibinfo {year}
  {2003})}\BibitemShut {NoStop}%
\bibitem [{\citenamefont {Holten}\ and\ \citenamefont
  {Anisimov}(2012{\natexlab{b}})}]{holtentwostatesupplement}%
  \BibitemOpen
  \bibfield  {author} {\bibinfo {author} {\bibfnamefont {V.}~\bibnamefont
  {Holten}}\ and\ \bibinfo {author} {\bibfnamefont {M.~A.}\ \bibnamefont
  {Anisimov}},\ }\href@noop {} {\bibfield  {journal} {\bibinfo  {journal} {Sci.
  Rep.}\ }\textbf {\bibinfo {volume} {2}},\ \bibinfo {pages} {713} (\bibinfo
  {year} {2012}{\natexlab{b}})},\ \bibinfo {note} {supplementary
  information}\BibitemShut {NoStop}%
\bibitem [{\citenamefont {Edison}, \citenamefont {Anisimov},\ and\
  \citenamefont {Sengers}(1998)}]{edison1998}%
  \BibitemOpen
  \bibfield  {author} {\bibinfo {author} {\bibfnamefont {T.~A.}\ \bibnamefont
  {Edison}}, \bibinfo {author} {\bibfnamefont {M.~A.}\ \bibnamefont
  {Anisimov}}, \ and\ \bibinfo {author} {\bibfnamefont {J.~V.}\ \bibnamefont
  {Sengers}},\ }\href@noop {} {\bibfield  {journal} {\bibinfo  {journal} {Fluid
  Phase Equilib.}\ }\textbf {\bibinfo {volume} {150--151}},\ \bibinfo {pages}
  {429} (\bibinfo {year} {1998})}\BibitemShut {NoStop}%
\bibitem [{\citenamefont {van~'t Hof}, \citenamefont {Japas},\ and\
  \citenamefont {Peters}(2001)}]{vanthof2001}%
  \BibitemOpen
  \bibfield  {author} {\bibinfo {author} {\bibfnamefont {A.}~\bibnamefont
  {van~'t Hof}}, \bibinfo {author} {\bibfnamefont {M.~L.}\ \bibnamefont
  {Japas}}, \ and\ \bibinfo {author} {\bibfnamefont {C.~J.}\ \bibnamefont
  {Peters}},\ }\href@noop {} {\bibfield  {journal} {\bibinfo  {journal} {Fluid
  Phase Equilib.}\ }\textbf {\bibinfo {volume} {192}},\ \bibinfo {pages} {27}
  (\bibinfo {year} {2001})}\BibitemShut {NoStop}%
\bibitem [{\citenamefont {Pelissetto}\ and\ \citenamefont
  {Vicari}(2002)}]{pelissetto2002}%
  \BibitemOpen
  \bibfield  {author} {\bibinfo {author} {\bibfnamefont {A.}~\bibnamefont
  {Pelissetto}}\ and\ \bibinfo {author} {\bibfnamefont {E.}~\bibnamefont
  {Vicari}},\ }\href@noop {} {\bibfield  {journal} {\bibinfo  {journal} {Phys.
  Rep.}\ }\textbf {\bibinfo {volume} {368}},\ \bibinfo {pages} {549} (\bibinfo
  {year} {2002})}\BibitemShut {NoStop}%
\bibitem [{\citenamefont {Sengers}\ and\ \citenamefont
  {Shanks}(2009)}]{sengers2009}%
  \BibitemOpen
  \bibfield  {author} {\bibinfo {author} {\bibfnamefont {J.~V.}\ \bibnamefont
  {Sengers}}\ and\ \bibinfo {author} {\bibfnamefont {J.~G.}\ \bibnamefont
  {Shanks}},\ }\href@noop {} {\bibfield  {journal} {\bibinfo  {journal}
  {J.~Stat. Phys.}\ }\textbf {\bibinfo {volume} {137}},\ \bibinfo {pages} {857}
  (\bibinfo {year} {2009})}\BibitemShut {NoStop}%
\bibitem [{\citenamefont {Povodyrev}, \citenamefont {Anisimov},\ and\
  \citenamefont {Sengers}(1999)}]{povodyrev1999}%
  \BibitemOpen
  \bibfield  {author} {\bibinfo {author} {\bibfnamefont {A.~A.}\ \bibnamefont
  {Povodyrev}}, \bibinfo {author} {\bibfnamefont {M.~A.}\ \bibnamefont
  {Anisimov}}, \ and\ \bibinfo {author} {\bibfnamefont {J.~V.}\ \bibnamefont
  {Sengers}},\ }\href@noop {} {\bibfield  {journal} {\bibinfo  {journal}
  {Physica A}\ }\textbf {\bibinfo {volume} {264}},\ \bibinfo {pages} {345}
  (\bibinfo {year} {1999})}\BibitemShut {NoStop}%
\bibitem [{\citenamefont {Poole}, \citenamefont {Saika-Voivod},\ and\
  \citenamefont {Sciortino}(2005)}]{poole2005}%
  \BibitemOpen
  \bibfield  {author} {\bibinfo {author} {\bibfnamefont {P.~H.}\ \bibnamefont
  {Poole}}, \bibinfo {author} {\bibfnamefont {I.}~\bibnamefont {Saika-Voivod}},
  \ and\ \bibinfo {author} {\bibfnamefont {F.}~\bibnamefont {Sciortino}},\
  }\href@noop {} {\bibfield  {journal} {\bibinfo  {journal} {J. Phys.: Condens.
  Matter}\ }\textbf {\bibinfo {volume} {17}},\ \bibinfo {pages} {L431}
  (\bibinfo {year} {2005})}\BibitemShut {NoStop}%
\bibitem [{\citenamefont {Tanaka}(1999)}]{tanaka1999}%
  \BibitemOpen
  \bibfield  {author} {\bibinfo {author} {\bibfnamefont {H.}~\bibnamefont
  {Tanaka}},\ }\href@noop {} {\bibfield  {journal} {\bibinfo  {journal} {J.
  Phys.: Condens. Matter}\ }\textbf {\bibinfo {volume} {11}},\ \bibinfo {pages}
  {L159} (\bibinfo {year} {1999})}\BibitemShut {NoStop}%
\bibitem [{\citenamefont {Biddle}\ \emph {et~al.}(2013)\citenamefont {Biddle},
  \citenamefont {Holten}, \citenamefont {Sengers},\ and\ \citenamefont
  {Anisimov}}]{biddle2013}%
  \BibitemOpen
  \bibfield  {author} {\bibinfo {author} {\bibfnamefont {J.~W.}\ \bibnamefont
  {Biddle}}, \bibinfo {author} {\bibfnamefont {V.}~\bibnamefont {Holten}},
  \bibinfo {author} {\bibfnamefont {J.~V.}\ \bibnamefont {Sengers}}, \ and\
  \bibinfo {author} {\bibfnamefont {M.~A.}\ \bibnamefont {Anisimov}},\
  }\href@noop {} {\bibfield  {journal} {\bibinfo  {journal} {Phys. Rev. E}\
  }\textbf {\bibinfo {volume} {87}},\ \bibinfo {pages} {042302} (\bibinfo
  {year} {2013})}\BibitemShut {NoStop}%
\bibitem [{\citenamefont {Flory}(1953)}]{florybook}%
  \BibitemOpen
  \bibfield  {author} {\bibinfo {author} {\bibfnamefont {P.~J.}\ \bibnamefont
  {Flory}},\ }\href@noop {} {\emph {\bibinfo {title} {Principles of Polymer
  Chemistry}}}\ (\bibinfo  {publisher} {Cornell University Press},\ \bibinfo
  {address} {Ithaca, NY},\ \bibinfo {year} {1953})\BibitemShut {NoStop}%
\end{thebibliography}%

\end{document}